\documentclass{aa}
\usepackage{graphicx}
\usepackage{natbib}

\begin{document}


\title{NGC 1866: a workbench for stellar evolution}

\author{R. Barmina, L. Girardi, and 
           C. Chiosi}
\institute{
     Department of Astronomy, University of Padova,
      Vicolo dell'Osservatorio 2, 35122 Padova, Italy   }

\offprints{Roberto Barmina \\ 
e-mail: barmina@usm.uni-muenchen.de}

\date{Received {August, 2001}. Accepted {} }

\authorrunning{Barmina et al.}
\titlerunning{NGC 1866: a workbench for stellar evolution}

\abstract{ 
NGC 1866 is a young, rich star cluster in the Large Magellanic Cloud. 
Since the cluster is  
very well populated both in the main sequence and post main sequence 
stages, thus  providing us with a statistically complete 
sample of objects throughout the various evolutionary phases of
intermediate mass stars,
it represents a good laboratory 
for testing stellar evolutionary models. 
More precisely, NGC 1866 can be used to discriminate among
classical stellar models, in which the extension of the 
convective regions is fixed 
by the classical Schwarzschild criterion, from models with 
overshooting, in which an ``extra-mixing'' is considered to take 
place beyond the classical limit of the convective zone. 
Addressing this subject in a recent work, \citet{Te99} reached the 
conclusion that the classical scheme for the treatment 
of convection represents a good and sufficient approximation for 
convective interiors.
Using their own data, we repeat here the analysis. First we revise 
the procedure followed by \citet{Te99} to correct the data for 
completeness,
second we calculate new stellar models with updated physical input
for both evolutionary schemes, finally we present many simulations
of the colour-magnitude diagrams and 
luminosity functions of the cluster using the ratio of the 
integrated luminosity 
function of  main sequence stars to the number of 
giants as the normalization factor of the simulations. We also take
into account several possible physical agents that could alter the
color-magnitude diagram and the luminosity function: they are unresolved
binary stars, dispersion in the age, stochastic effects in the initial
 mass function. Their effect is analyzed separately, with the conclusion 
that binary stars have the largest impact. 
The main result of this study is that 
the convective overshoot hypothesis (together with a suitable 
percentage of 
unresolved binaries) is really needed to fully match the whole 
pattern of data.  The main drawback of the 
classical models is that they cannot reproduce the correct ratio of 
main sequence  to post-main sequence stars. 
}
\maketitle

\section{Introduction}\label{introd}
In the context of stellar theories, it has long been acknowledged that the 
penetration of convective elements beyond the classical limit set by 
the Schwarzschild criterion could produce non-negligible effects on the 
stellar structure.

The young star cluster NGC 1866 in the Large Magellanic Cloud (LMC)
is considered as an ideal laboratory for testing  stellar models, 
especially concerning the  extension of convection in the 
interiors of real stars, and discriminating among classical and so-called 
overshooting models.  
This cluster, well populated both in main sequence and giants 
stars, represents a statistically complete sample of objects, 
contrary to the young clusters of the Milky Way (that are 
relatively poor, especially in 
giant stars). Furthermore, NGC 1866 is young enough ($\simeq$100 Myr) to 
possess stars which 
develop a convective core in the phase of central H-burning 
($M_{\rm TO}=4 - 5$ M$_{\odot}$), thus providing us the way 
to test the extension of overshooting.  The theory of stellar structure
predicts that convective cores in main sequence stars set in starting
from initial masses greater  than 
$M \simeq 1.1$ M$_{\odot}$. 
For these reasons NGC 1866 has been the subject of several analyses, 
aiming at testing how far convective elements overshoot from the core 
into the surrounding stable layers.

\citet{BM83}, comparing observational and 
synthetic colour-magnitude diagrams (CMD) of NGC 1866, obtained the 
best fit to  the data simulating a cluster of about $86$ Myr: 
the general features of the CMD were well-fitted on the whole, but the 
authors  argued that the classical stellar models they used predicted
a  number of red giants larger than  observed, and a smaller number of 
main sequence stars in turn as compared to the observations.
The same authors suggested that a more careful treatment of core convection
(i.e. larger convective cores) could remove the discrepancy. 
A number of subsequent studies confirmed this idea and demonstrated 
that the inclusion of overshooting in the description of convective motions
could reproduce the correct ratio $N_{\rm PMS}/N_{\rm MS}$.
\citet{C89a}, in particular,  clearly showed that the overshooting scheme, 
by reducing the ratio $t_{\rm He}/t_{\rm H}$, offers a good and simple 
solution to the problem. This conclusion was also reinforced by 
\citet{L91}, who analyzed the CMD of NGC 1866 by means of overshooting and
classical models \footnote{It is worth recalling here that models of
intermediate mass stars calculated with the
classical Schwarzschild criterion are known to develop during 
the core He-burning
phase the so-called He-semi-convective instability, i.e. a region surrounding
the fully convective core in which the condition of neutrality 
$\nabla_R=\nabla_A$ is maintained by suitably modifying the profile of
chemical composition. This is the physical analog of what happens in massive
stars during the core H-burning phase, which develop the so-called
H-semi-convective instability. More details on the physical 
origins of both H- and He-semi-convection,
can be found in \citet{C92}. Let it suffice here to mention
that in intermediate-mass stars 
the effects of semiconvection are negligible, if compared to the effects
of convective overshooting. Hereinafter this type of
stellar models are
referred to as the classical, semi-convective models.}.

Over the years, 
sophisticated formulations of convective overshooting have been elaborated 
that are based on turbulence theories \citep{Cloutman80,Xiong80} and 
fluid hydro-dynamics \citep{Canuto91,Canuto96,UnnoKondo89}. However, the
ballistic approach at the problem proposed by 
\citet{Br81} turns out to be fully adequate and it has been
proved to best reproduce the 
numerical results of laboratory fluid-dynamics simulations 
\citep{Zahn91}. The \citet{Br81} algorithm adopts a 
non-local treatment of convection in the context of the mixing-length 
theory (MLT)  by  \citet{Bohm55}: it looks for the layers where the
velocity of convective elements (accelerated by the buoyancy force 
in the formally unstable regions) becomes zero in the surrounding 
stable regions, then adopts a suitable temperature stratification in the
overshooting regions,  and finally assumes straight mixing over-there.
Since the \citet{Br81} formalism makes use of the MLT, it expresses 
the mean free 
path of the convective elements as  $l=\Lambda_{\rm c} \times H_{\rm p}$ where
 $H_{\rm p}$ is the local pressure scale height.

Although some criticism has been advanced by \citet{Renzini87} 
--who erroneously 
concludes that the ``miscellaneous'' of local and non local 
formalisms leads to overestimating the  overshooting distance-- we remind 
the reader that 
overshooting is simply a logical consequence of the inertia principle, 
so that neglecting its existence would not be physically sound.
 It is worth recalling that convective 
 overshooting is quite common 
in nature \citep{Deardorff69}, and it has been demonstrated 
in a number of studies, including numerical simulations 
\citep{Freytag96}, that 
the penetration depth of convective elements into a formally 
stable region represents a non-negligible fraction of the size of 
the unstable zone.

In addition to this, there are a number of astrophysical situations
in which the hypothesis of substantial convective overshooting 
has been found to offer better and more elegant solutions than 
other explanations 
\citep[see][]{Be86,C92}. Among others, it suffices to recall here
the so-called mass discrepancy of Cepheid stars \citep{Bw93,C92}.

Despite this, it has been often argued that 
 unresolved binaries could mimic the effects of convective overshooting
 and easily  
account for the low ratio of red giant to main sequence stars 
 observed in the young LMC  clusters. 
 This hypothesis has been investigated by many authors, 
both in open clusters of the Milky Way \citep[see][]{Ca94} and 
young clusters of the 
LMC \citep{C89a,C89b,VaC92} with somewhat contrasting results. 

In their study, \citet{Te99} emphasize the role
of unresolved 
binaries in solving the problem, reaching  
the conclusion that overshooting is not needed.  They 
obtain agreement with the observational data only by 
introducing a fraction of about $30\,\%$ of binary stars 
in a population of $\simeq 100$ Myr, 
and using the classical semi-convective models of \citet{Dominguez99}. 
In contrast, models computed with ``enlarged'' convective cores 
(i.e. cores in which convection extends beyond 
the Schwarzschild criterion to simulate overshooting), 
lead to worse fits of the observational  data, especially when 
 binary stars are included.

Unfortunately,  \citet{Te99} adopt a  
wrong procedure to correct the observational star counts  for photometric 
completeness  (see the section below), 
thus obtaining differential ($N_{\rm MS}$, shortly indicated with 
DLF) and integrated luminosity functions 
($\Sigma N_{\rm MS}$, shortly referred to as ILF) for the 
main sequence stars of NGC 1866 that are inconsistent.
Furthermore, \citet{Te99} fail to make use of the ILF  normalized 
to the number of evolved stars ($\Sigma N_{\rm MS}/N_{\rm PMS}$, 
hereinafter indicated as N-ILF)    introduced by \citet{C89a}, 
which has been proved 
to be the only way to effectively discriminate 
between the two different evolutionary schemes. It is worth reminding the 
reader that the N-ILF is by definition proportional to the lifetime ratio
$t_{\rm H}/t_{\rm He}$.     

In this paper, first we correct the analysis of the observational data
and second we show that the conclusions reached by \citet{Te99} entirely
depend on the kind of diagnostic they have adopted. Exploring 
the relative effects of core overshoot and unresolved
binary star frequency on intermediate age cluster CMD's will be 
a side product of our analysis.

Starting from the same observational data used by 
\cite{Te99}\footnote{All the photometric data used in this study were 
kindly provided by \cite{Te99} already reduced and calibrated.}, 
and correctly applying the completeness 
correction, in this paper we shall discriminate --by means 
of simulated CMDs and LFs of 
NGC 1866-- between classical semi-convective  and  
 overshooting models.  
After a brief description of the observational data  (Sect.
\ref{sec:data}), in Sect. \ref{sec:tracks} we 
compare the stellar models calculated 
according to the two schemes, with particular attention to
the values of critical masses and lifetime ratios.  These latter are 
indeed the crucial point to test the validity of stellar models. 
In Sect. 
\ref{sec:isochrones} we describe the isochrones  and the methods used 
to simulate CMDs and LFs. In Sect. 
\ref{sec:simulations} we compare the results of these simulations with 
their observational counterparts. In Section \ref{why:testa} we 
thoroughly
discuss the reasons why our conclusions differ from those by \citet{Te99}.
Finally, in Sect. \ref{sec:conclusions} 
we summarize the main results of this study.

\begin{figure}
\resizebox{\hsize}{!}{\includegraphics{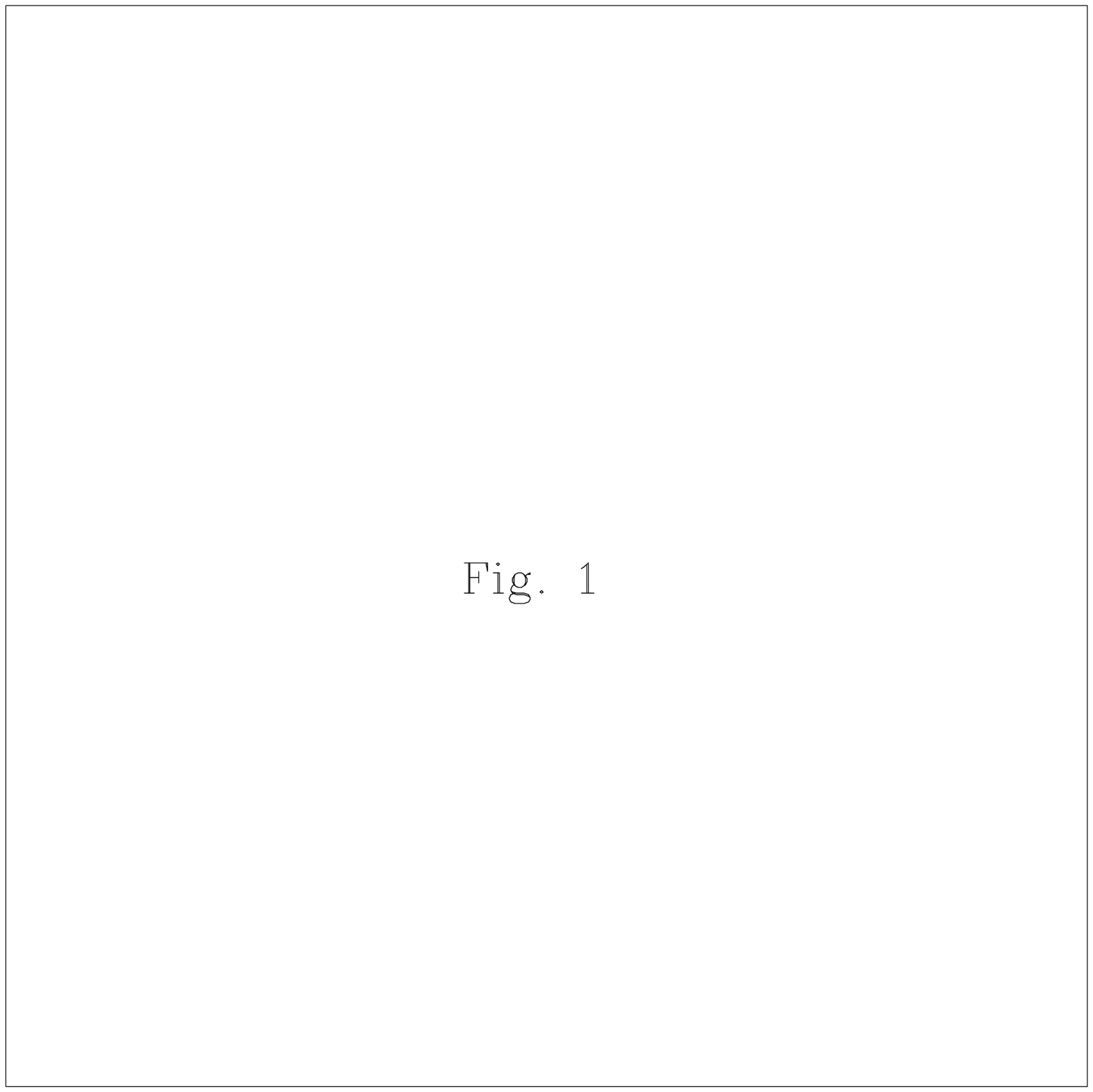}}
\caption{CMD for stars in the adopted sample, 
	after statistical de-contamination \citep{Te99}}
	\label{fig:1866_grid}
\end{figure}

\section{Observational data}\label{sec:data}

The observational data \citep{Te99} have been obtained with the 
$2.2$ m ESO telescope at La Silla, Chile, on 1993 January 27. Two fields 
have been observed in $B$ and $V$ pass-bands: the  
 first field is centered on the cluster, and the second one is  
located at $\sim 6^\prime$ from it. The second field is
utilized  for statistical field subtraction: the corrected and calibrated 
CMD  of \citet{Te99} is presented in Fig. \ref{fig:1866_grid}, in which 
 the main sequence stars are separated from  
giants by dashed lines.

\begin{table*}
\caption{\small Luminosity functions for main sequence stars 
	in NGC~1866 as adopted in \citet{Te99}: differential and
integrated 
	luminosity functions before the completeness 
	correction (columns 2 ad 3), the completeness factors (column 4), 
	differential and integrated luminosity functions after the
completeness 
	correction (column 5 and 6).  
	The last three columns present the corrected 
luminosity functions as obtained independently by us, starting from the
original \citet{Te99} data.}
\label{tab:factors}
\begin{center}
\begin{tabular}{c|rrcrr|rrr}
\hline\noalign{\smallskip}
  \multicolumn{1}{c|}{} & \multicolumn{5}{c|}{ \citet{Te99}} &
\multicolumn{3}{c}{ This work} \\
\hline\noalign{\smallskip}
(1)  &  (2) &  (3) &   (4) &  (5) &  (6) & (7) &  (8) & (9)  \\
\noalign{\smallskip}\hline\noalign{\smallskip}
$m_{\rm V}$ & $N_{\rm MS}$ & $\Sigma N_{\rm MS}$ & $\lambda_{\rm c}$ 
& $(N_{\rm MS})_{\rm corr}$ & $\Sigma (N_{\rm MS})_{\rm corr}$
& $N_{\rm MS}$ 
& $(N_{\rm MS})_{\rm corr}$ & $\Sigma (N_{\rm MS})_{\rm corr}$\\
\noalign{\smallskip}\hline\noalign{\smallskip}
16.5 &    0 &    0 & 1.000 &    0 &    0 &   3  &   3 &    3 \\
16.7 &    2 &    2 & 1.000 &    2 &    2 &   3  &   3 &    6 \\ 
16.9 &    9 &   11 & 1.000 &    9 &   11 &  10  &  10 &   16 \\
17.1 &   17 &   28 & 1.000 &   17 &   28 &  20  &  20 &   36 \\
17.3 &   37 &   65 & 1.000 &   37 &   65 &  37  &  37 &   73 \\
17.5 &   29 &   94 & 1.000 &   29 &   94 &  33  &  33 &  106 \\
17.7 &   56 &  150 & 1.000 &   56 &  150 &  59  &  59 &  165 \\
17.9 &   55 &  205 & 1.000 &   55 &  205 &  61  &  62 &  227 \\
18.1 &   69 &  274 & 0.996 &   69 &  275 &  74  &  74 &  301 \\
18.3 &   90 &  364 & 0.992 &   90 &  367 &  84  &  85 &  386 \\
18.5 &  116 &  480 & 0.988 &  117 &  486 & 129  & 131 &  517 \\
18.7 &  135 &  615 & 0.981 &  137 &  627 & 144  & 146 &  663 \\
18.9 &  160 &  775 & 0.975 &  163 &  795 & 167  & 172 &  835 \\
19.1 &  164 &  939 & 0.966 &  168 &  972 & 174  & 180 & 1015 \\
19.3 &  188 & 1127 & 0.958 &  194 & 1177 & 208  & 217 & 1232 \\
19.5 &  215 & 1342 & 0.950 &  222 & 1413 & 227  & 238 & 1470 \\
19.7 &  266 & 1608 & 0.939 &  279 & 1712 & 276  & 292 & 1762 \\
19.9 &  250 & 1858 & 0.929 &  263 & 2001 & 256  & 274 & 2036 \\
20.1 &  274 & 2132 & 0.916 &  292 & 2327 & 274  & 297 & 2333 \\
20.3 &  323 & 2455 & 0.903 &  352 & 2720 & 313  & 346 & 2679 \\
20.5 &  303 & 2758 & 0.889 &  332 & 3101 & 316  & 351 & 3030 \\
20.7 &  322 & 3080 & 0.874 &  355 & 3524 & 299  & 334 & 3364 \\
20.9 &  302 & 3382 & 0.859 &  336 & 3939 & 295  & 331 & 3695 \\
21.1 &  290 & 3672 & 0.842 &  323 & 4362 & 282  & 318 & 4013 \\
21.3 &  316 & 3988 & 0.823 &  358 & 4843 & 289  & 332 & 4345 \\
21.5 &  261 & 4249 & 0.803 &  299 & 5290 & 238  & 278 & 4623 \\
21.7 &   86 & 4335 & 0.776 &  100 & 5587 &  88  & 106 & 4729 \\
21.9 &   16 & 4351 & 0.744 &   29 & 5850 &  36  &  50 & 4779 \\
\noalign{\smallskip}\hline\noalign{\smallskip}
\end{tabular}
\end{center}
\end{table*}

\subsection{Correction for completeness}

In their analysis, \citet{Te99} divide the sample in six 
concentric rings centered on the cluster,  
and calculate the completeness factors, $\lambda^i_{\rm c}$, 
ring by ring. Since $\lambda^i_{\rm c}$ is equal to the number
ratio of artificial stars added and recovered in a given 
magnitude bin, they should be used to correct the
DLF -- i.e.\ the number of stars per 
magnitude bin -- separately for each ring $i$:
	\begin{equation}
(N^i)_{\rm corr} = N^i / \lambda^i_{\rm c}
	\label{eq:compl}
	\end{equation}
where $N^i$ and $(N^i)_{\rm corr}$ stand for the observed
and corrected numbers, respectively.

However, a substantial error is made by \citet{Te99}:
they use the same $\lambda_{\rm c}$ factors to 
{\em correct both the number of stars 
and the ILF at each magnitude bin}. This 
clearly produces the correct DLF $(N^i)_{\rm corr}$, but
 over-estimates  the ILF and N-ILF
at fainter magnitude bins.
This can be appreciated by looking at the first 6 columns
of Table \ref{tab:factors}, where we
present the original star counts and the  luminosity functions used by 
\citet{Te99}\footnote{These data have been kindly provided by 
V.\ Testa.}. In particular, it can be noticed that, starting from
the uncorrected DLF and ILF of columns 2 and 3, the same ``average''
$\lambda_{\rm c}$ factors (column 4) have been used to derive
the corrected DLF shown in column 5 (i.e.\ 
$(N_{\rm MS})_{\rm corr} = N_{\rm MS} / \lambda_{\rm c}$), and the 
corrected ILF shown in column 6 (i.e.\ 
$\Sigma(N_{\rm MS})_{\rm corr} = \Sigma N_{\rm MS} / \lambda_{\rm c}$).
The corrected DLF and ILF obtained in such a way are no longer
consistent with each other.

Noticing this, 
we decided to completely re-derive the luminosity functions
for NGC~1866,
starting from the statistically decontaminated
photometric data of \citet{Te99} (see 
Fig.~\ref{fig:1866_grid}), and  the completeness factors presented in 
their Fig.~6. In short, we have obtained the 
DLF for each annulus around NGC~1866, 
and corrected them using the corresponding completeness factors 
separately, according to Eq.~(\ref{eq:compl}). 
Then, the DLFs for annulus 2 to 5
were added
	\begin{equation}
(N_{\rm MS})_{\rm corr} = \sum_{i=2}^5 (N^i_{\rm MS})_{\rm corr}
	\end{equation}
producing the total DLF presented in column 8 of 
Table \ref{tab:factors}. (For the sake of comparison, column 7 of the
same table presents the counts obtained by adding stars from 
annulus 2 to 5, without applying the completeness correction.) 
>From this corrected DLF we derive the total ILF of
column 9 of the same table, by simply summing up the number of stars
above each magnitude bin. As can be readily seen, our final
numbers for the DLF are roughly consistent with those of
\citet{Te99}, but the two ILFs are sizeably different, especially for
the faintest magnitude bins. 

There is a final remark to be made for the sake of clarity. Comparing
columns (2) and (7), which contain the rough counts in the various magnitude
bins, a marginal disagreement is evident: our counts 
do not
exactly coincide with those by \citet{Te99}. The difference is due to the
slightly different criteria adopted to define the numbers $N_{MS}$ of stars 
per  magnitude
interval. \citet{Te99} follow a complicated scheme in which (i) the mean
location of the MS together with its color dispersion $\sigma$ 
are derived as a 
function of the magnitude, and (ii) 
in every magnitude bin only stars  falling within 7$\sigma$ from the 
mean MS are taken into account. 
In contrast, we prefer to consider all stars falling within the
considered magnitude interval. As a consequence of this, the $N_{MS}$ of
\citet{Te99} are slightly smaller than ours because stars far away from 
the  mean MS are neglected.

\begin{figure}
\resizebox{\hsize}{!}{\includegraphics{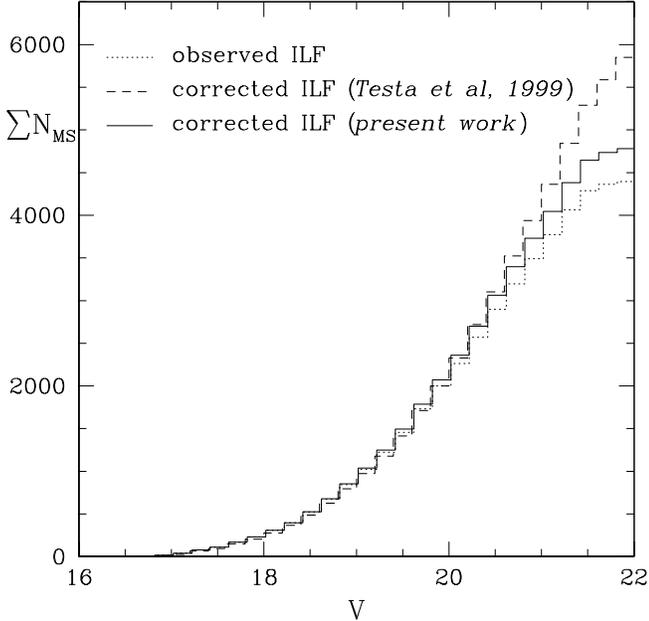}}
\caption{Integrated Luminosity Function of the  main sequence stars of 
NGC 1866 before (dotted lines) and after (thick line) correcting
for photometric completeness. The dashed line shows the 
same luminosity function calculated by   \citet{Te99}}
\label{fig:conf_corrections}
\end{figure}

In Fig. \ref{fig:conf_corrections} we compare our results
for the corrected ILF (continuous line) with those obtained 
by \citet{Te99} (dashed line). It is important to stress here that
one of  reasons why our results differ from those by \citet{Te99},
resides in the different (and erroneous) way of applying the
completeness corrections used by them.

Finally, we summarize in Table \ref{tab:counts} the total 
number of main sequence 
($\Sigma N_{\rm MS})$ and post main sequence ($N_{\rm PMS})$ stars 
before and after 
the completeness correction is applied.
The number of post main sequence stars amounts to $N_{\rm PMS}=100$.

\section{Stellar tracks}\label{sec:tracks}
We have calculated two sets of stellar tracks with the classical 
semi-convective scheme and 
initial chemical composition [$Y$=0.240,$Z$=0.004] and 
[$Y$=0.250,$Z$=0.008], 
and one set of models with convective overshooting and composition 
[$Y$=0.240,$Z$=0.004]. The initial masses of the models range 
from $1.0$ to $8$ M$_{\odot}$. The input physics is the 
same as in \citet{Gi00} and subsequently updated by \citet{Sa00}, to whom 
the reader should refer for all details. 
For purposes of comparison, whenever required we also
 utilize the stellar models by \citet{Sa00} 
calculated with overshooting and chemical composition 
[$Y$=0.250,$Z$=0.008].

\begin{table}
\caption{\small Star counts for main sequence ($\Sigma N_{\rm MS}$) 
	and post main sequence stars ($N_{\rm PMS}$), before and after 
	completeness correction.}\label{tab:counts}
\begin{center}
\begin{tabular}{l|cc}
\hline\noalign{\smallskip}
                           & $\Sigma N_{\rm MS}$  & $N_{\rm PMS}$\\
\noalign{\smallskip}\hline\noalign{\smallskip}
observed              &    4395     &    100\\
corrected             &    4779     &    100\\
\noalign{\smallskip}\hline
\end{tabular}
\end{center}
\end{table}

\subsection{Prescriptions for semi-convection and overshooting}

The algorithm dealing with semi-convection in classical models
strictly follows the recipes
given by \citet{Al93} to whom the reader should refer for all details.

Convective overshooting is based on the formulation by \citet{Br81} and the
more recent revision by \citet{Be85} and \citet{Al93}. Suffice it to recall
here that the overshooting parameter $\Lambda$ is chosen to be
$\Lambda_{\rm c}$=0.5 for core convection and 
$\Lambda_{\rm e}$=0.7 for envelope
convection   \citep[see][]{Al93}.

For the sake of clarity and to avoid misunderstanding, we remind the 
reader that  $\Lambda_{\rm c}$=0.5 in the formalism of  \citet{Br81} is
equivalent to  $\Lambda_{\rm c}$=0.25 in the  description of 
\citet{MaederMey91}.
More precisely, while  \citet{Br81} define the overshooting 
distance as the path 
traveled by a convective element across the Schwarzschild border (half 
distance beneath and half above the border),   \citet{MaederMey91}
define it only as the distance above the border.
Furthermore, the choice of $\Lambda_{\rm c}$=0.5 
we have adopted is the same as in previous  studies aimed at 
generating stellar models best suited to interpret observational data of
star clusters. 
To mention a few we recall \citet{Al93}, \citet{Br93a},
\citet{MaederMey91}, \citet{Meynet94}, \citet{C92}, \citet{Chiosi99} and
references therein. Exploring the effects of core overshooting for other 
choices of $\Lambda_{\rm c}$ is beyond the scope of this study.

\begin{figure}
\resizebox{\hsize}{!}{\includegraphics{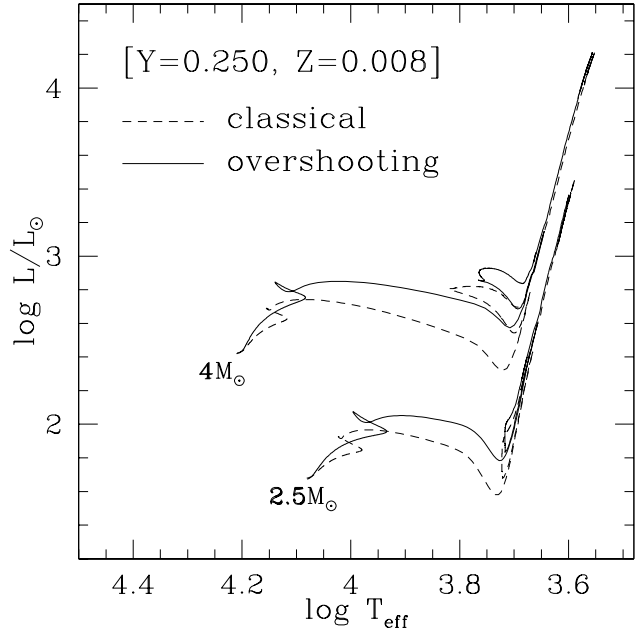}}
\caption{\small The stellar tracks for the $2.5$ e $4$ M$_{\odot}$ stars 
with composition [$Y$=0.250, $Z$=0.008]. The solid lines are the stellar
models with convective overshooting, whereas the dotted lines are the 
 models with the classical extension of the convective core}
\label{fig:conf_tracks}
\end{figure}

\subsection{Critical masses $M_{\rm HeF}$, $M_{\rm up}$}
As long ago pointed out by
\citet{C89a}, see also \citet{C92}, in the presence of core overshooting
 the minimum initial masses, 
below which core He-flash 
and core C-ignition in highly degenerate material occur, $M_{\rm HeF}$ 
and $M_{\rm up}$ respectively, get smaller. 
Table \ref{tab:critical_masses} summarizes the results obtained 
for  classical semi-convective and overshooting models with metallicity 
$Z$=0.004 and $Z$=0.008. The same result holds good for both values of 
$Z$.

\begin{table}
\caption{Critical masses $M_{\rm HeF}$ e $M_{\rm up}$ --
	in units of M$_{\odot}$ -- for classical semi-convective and   
	 overshooting models.}	
	\label{tab:critical_masses}  
\begin{center}                 
\begin{tabular}{c | c  c}
\hline\noalign{\smallskip}
& $M_{\rm HeF}$ & $M_{\rm up}$\\
\noalign{\smallskip}\hline\noalign{\smallskip}
classical          &  $2.2 - 2.5$ & $>8$\\
overshooting       &  $1.8 - 2.0$   & $\simeq 7$\\
\noalign{\smallskip}\hline
\end{tabular}
\end{center}
\end{table}

\begin{table}
\caption{Lifetimes for classical models with chemical composition
[$Y$=0.250,$Z$=0.008]. The lifetimes are in Gyr.}
	\label{tab:classic_times}
\begin{center}                      
\begin{tabular}{c c c c}
\hline\noalign{\smallskip}
$M/M_{\odot}$ & $t_{\rm H}$  &  $t_{\rm He}$ & $t_{\rm He}/t_{\rm H}$\\
\noalign{\smallskip}\hline\noalign{\smallskip}
2.5  & 0.4094  & 0.2436  & 0.595\\
3.5  & 0.1769  & 0.1293  & 0.389\\
4.0  & 0.1271  & 0.0688  & 0.340\\
5.0  & 0.0775  & 0.0219  & 0.283\\
6.0  & 0.0524  & 0.0126  & 0.241\\
7.0  & 0.0385  & 0.0079  & 0.208\\
8.0  & 0.0299  & 0.0056  & 0.189\\
\noalign{\smallskip}\hline
\end{tabular}
\end{center}
\end{table}

\begin{table}
\caption{The same as in Table~\protect\ref{tab:classic_times} but for the 
  overshooting models.}
	\label{tab:overs_times}
\begin{center}                      
\begin{tabular}{c c c c}
\hline\noalign{\smallskip}
$M/M_{\odot}$ & $t_{\rm H}$  &  $t_{\rm He}$ & $t_{\rm He}/t_{\rm H}$\\
\noalign{\smallskip}\hline\noalign{\smallskip}
2.5  & 0.5438  & 0.1376  & 0.253\\
3.5  & 0.2308  & 0.0380  & 0.165\\
4.0  & 0.1674  & 0.0237  & 0.141\\
5.0  & 0.0999  & 0.0116  & 0.116\\
6.0  & 0.0672  & 0.0065  & 0.096\\
7.0  & 0.0487  & 0.0043  & 0.090\\
8.0  & 0.0376  & 0.0032  & 0.085\\
\noalign{\smallskip}\hline
\end{tabular}
\end{center}
\end{table}

\subsection{Lifetimes and lifetime ratios}

Because of the increased size of the convective cores, the central 
H-burning 
lifetime $t_{\rm H}$ in the overshooting  models is  $30\,\%$ longer than in 
classical ones, this  
effect depending on the parameter $\Lambda_{\rm c}$ \citep{Br81}. The 
``over-luminosity'' caused by overshooting during the core H-burning 
phase (see Fig. \ref{fig:conf_tracks})
still remains during the He-burning phase (for stars with initial mass 
greater 
than $M_{\rm HeF}$): consequently, the lifetime of the He-burning phase  
($t_{\rm He}$) gets shorter by about $40 - 45\,\%$. This, combined with 
the 
longer $t_{\rm H}$, results in a strong decrease ($\sim 55 - 60\,\%$) 
of the ratio $t_{\rm He} / t_{\rm H}$.
Tables \ref{tab:classic_times} and \ref{tab:overs_times} summarize the 
lifetimes for some of the $Z$=0.008 models calculated with the classical 
and overshooting scheme, respectively.

\section{Isochrones and synthetic CMDs}\label{sec:isochrones}

\subsection{Isochrones}
For the above four groups of evolutionary tracks, we 
have derived the  isochrones, adopting the algorithm 
of ``equivalent evolutionary points'' developed by  \citet{Be94}. The age 
ranges from about $0.05$ to $10$ Gyr.
Isochrones are calculated at $\Delta \log t=0.01$ intervals: this means 
that any two consecutive isochrones differ by $\sim 2\,\%$ in their ages. 
Along  each isochrone we give: the initial and actual stellar 
masses, the logarithm of surface luminosity, the 
effective temperature and surface 
gravity, the absolute bolometric magnitude, and the absolute magnitudes 
in the $UBVRIJHK$ pass-bands. Finally we list the indefinite integral 
over the initial mass $M$ of the initial mass function (IMF). For any 
other detail see  \citet{Gi00}.

Theoretical luminosities and effective temperatures are translated into 
magnitudes and colours in the Johnson-Cousins  system, using the 
conversions of \citet{Be94},  which provides colours and 
bolometric corrections as a function of the effective temperature and 
gravity. The \citet{Be94} conversions are based on the 
library of synthetic spectra obtained 
from theoretical models of stellar atmospheres by  \citet{Ku92}.

\begin{figure}
\resizebox{\hsize}{!}{\includegraphics{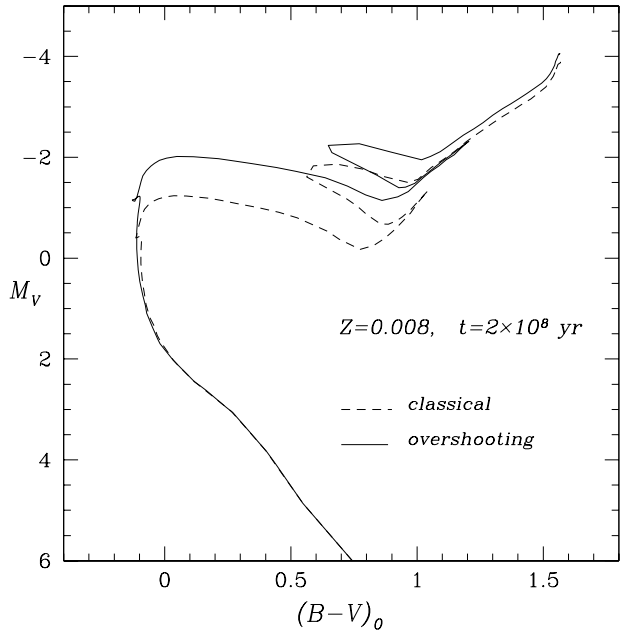}}
\caption{Isochrones  of same age ($200$ Myr) and chemical composition,
[$Y$=0.250,$Z$=0.008] calculated with classical  
(dashed line) and  overshooting 
models (solid line)}\label{fig:iso_conf_cl_ov}
\end{figure}

We only note here, as illustrated in Fig. \ref{fig:iso_conf_cl_ov}, that 
an isochrone of the  overshooting scheme runs at 
higher luminosities as compared to a ``classical'' isochrone of the same 
age, that is, in the presence of  overshooting   isochrones 
of older ages are required to fit the observational data of a cluster.

\subsection{Synthetic CMDs}
The synthetic CMDs  are constructed by means of a 
Monte Carlo algorithm, which randomly distributes stars along a 
given isochrone according 
to evolutionary lifetimes and IMF. The code takes into account the 
photometric errors affecting the  observational data
(derived from \citet{Te99} tables), by adding an artificial 
dispersion to the theoretical magnitudes and colours, and generates
numerical simulations of CMDs allowing for effects due to: 
age spread of the cluster population, different slopes of the IMF, and  
presence of a certain fraction of unresolved binary stars with mass 
ratio ranging in a given interval. 

Since any plausible IMF preferentially populates the main sequence 
region of an isochrone, in order to reproduce the right proportion 
of main sequence to post main sequence stars --according 
to the lifetimes in the various evolutionary phases-- we populate our 
synthetic CMDs until an assigned number of post main 
sequence stars $N_{\rm PMS}$ is matched: in particular 
in each simulation presented here, we impose that 
the total number of red giant stars present in the data is matched. In 
such a way $N_{\rm PMS}$ is taken as the ``normalization'' parameter 
of the simulations \citep[see][]{C89a,V91,L91}.

\subsection{Luminosity functions}
Special care is paid to the integrated number of stars from 
the tip of main sequence band down to the current magnitude 
interval, normalized to the number of red giant stars (i.e. the
N-ILF).
As pointed out by 
\citet{C89a}, the N-ILF 
can be directly compared with the ratio of lifetimes in the core H- and 
He-burning phases, $t_{\rm H}$ and $t_{\rm He}$, respectively. 
At fixed $N_{\rm PMS}$, the ratio in each simulation 
$\Sigma N_{\rm MS}/N_{\rm PMS}$ depends on the age 
--an older population clearly has a larger $N_{\rm PMS}$-- and  
the theoretical model in use. As already 
noted in  previous sections, the size of the convective cores strongly 
affects the ratio $t_{\rm He}/t_{\rm H}$, and in turn, the ratio 
$\Sigma N_{\rm MS}/N_{\rm PMS}$. Thus the N-ILF, when compared to the 
observational counterpart,  reliably discriminates  classical from
 overshooting models. 

Concluding this section, it is worth remarking that reproducing  
synthetic CMDs with an assigned number of post main sequence stars 
is by far  preferable to  simulating 
CMDs with a fixed total number of stars brighter than 
a certain magnitude. As a matter of  fact, the 
distribution of stars along the main sequence is driven almost 
exclusively by 
the IMF,  evolutionary effects playing only a marginal role. 
Therefore, 
the LF  (either DLF or ILF) of the main sequence stars,
if not normalized to the number of giants,
would only reflect the underlying 
IMF, and would not be  
affected by the effects of overshooting  we want to test.

\section{The simulations}\label{sec:simulations}
In this section we present the CMDs and LFs resulting 
from our simulations, and compare them with the observational ones,  both 
for classical semi-convective and overshooting models.

\subsection{Adopted parameters}
Before presenting the simulations, we briefly summarize the values 
adopted for parameters such as reddening, metallicity, and distance 
modulus.

\subsubsection{IMF}
All the simulations are calculated  adopting the \citet{Salp55} law 
$dN \propto M^{- \alpha}dM$ for the IMF and assuming, 
unless otherwise specified, the parameter $\alpha =2.35$.

\subsubsection{Metallicity}
Rough estimates of the metallicity of NGC 1866 derived from the 
$BVI_{\rm c}$ colors of the Cepheids  \citep{Cald85,Feast89} yield 
[Fe/H]$=-0.1\pm0.3$
\citep[see also ][]{Bw93}, finally  the recent analysis 
by \citet{Te99} yielded [Fe/H]$\simeq -0.35$. This  roughly corresponds to 
$Z=0.009$. Therefore, improving upon the metallicity used in the old
studies by \citet{C89a} and \citet{Broca89} who adopted 
isochrones with [$Y$=0.280,$Z$=0.020] in the present simulations we
assume the composition  [$Y$=0.250,$Z$=0.008].

\subsubsection{Distance modulus}
The distance modulus to the LMC is a hotly debated issue: we   summarize 
here some of the most recent determinations of this parameter. 
Cepheid stars with {\it Hipparcos} parallaxes indicate
$(m-M)_{\rm 0}\sim18.6$ \citep{Feast00}. 
The analysis of the expanding ring around 
 SN1987A yields $(m-M)_{\rm 0}=18.58 \pm 0.05$ \citep{Panagia98}.
The simultaneous study of the Cepheids stars and CMD of NGC 1866 by 
\citet{Bw93}
yields  $(m-M)_{\rm 0}=18.51\pm0.21$
In a recent study on young LMC clusters \citet{Keller00} utilize the
value  $(m-M)_{\rm 0}=18.45$.
We adopt here $(m-M)_{\rm 0}=18.5$, as the  
``classical'' value for the LMC distance modulus \citep{Westerlund97}.
Passing from $(m-M)_{\rm 0}=18.5$ to $(m-M)_{\rm 0}=18.6$ would not
significantly change the  results of the present analysis. 

\subsubsection{Reddening}
The study of the dust distribution across the LMC by means of 
photometric and 
spectroscopic data gives reddening maps from which one derives an
estimate of the mean value $E_{B-V}=0.16$ \citep{Oes96}, with a maximum 
value of 0.29 (reached in the region of {\it 30 Dor}) and a minimum of 
0.06 \citep{Z97}. While \citet{Te99} 
assume $E_{B-V}$ in the range between $0.06$ to $0.10$, we prefer to
adopt the value 
$0.10$ which, taking the total visual extinction $A_{V}$ to be 
$3.1\times E_{B-V}$ \citep{Rieke85}, corresponds to $A_{V}=0.31$.

\subsection{Classical semi-convective models}
\begin{figure}
\resizebox{\hsize}{!}{\includegraphics{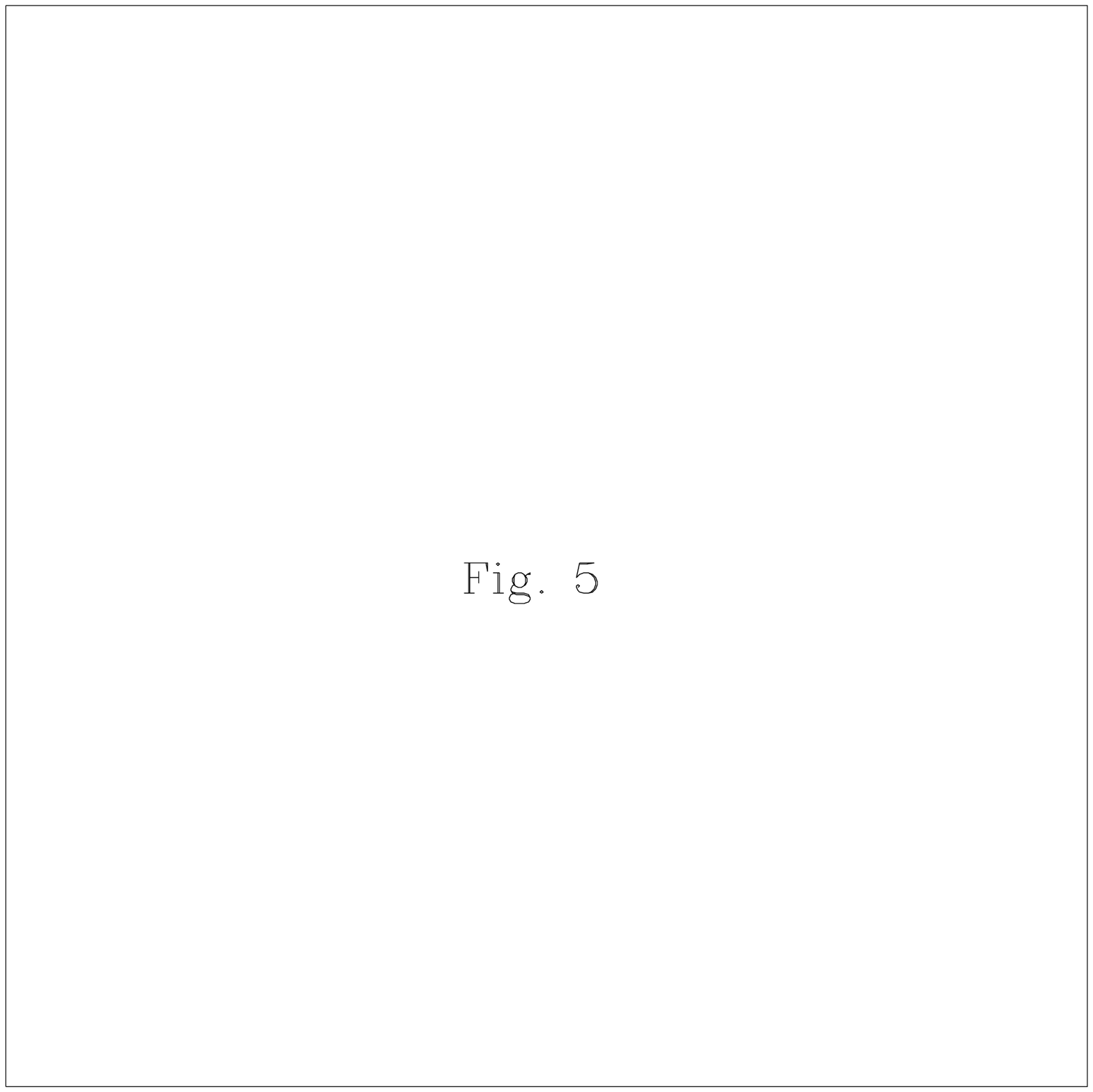}}
\caption{\small Synthetic CMD constructed with 
classical 
	models. The age is  $76$ Myr,  the  chemical composition is 
	[$Y$=0.250,$Z$=0.008].}\label{fig:classic_single_CM}
\end{figure}

\begin{figure}
\resizebox{\hsize}{!}{\includegraphics{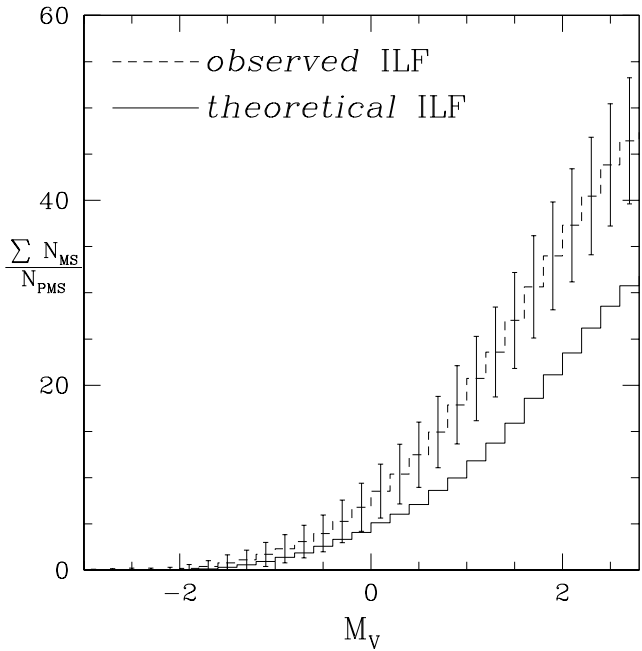}}
\caption{\small Observational (dashed line) N-ILF ---with the poissonian 
error bars--- and theoretical N-ILF (thick line) for the 
case of classical models shown in  
Fig. \protect{\ref{fig:classic_single_CM}}. The age is
of $76$ Myr and the chemical composition is [$Y$=0.250,$Z$=0.008]
}\label{fig:classic_single_ILF}
\end{figure}

In Fig. \ref{fig:classic_single_CM} 
is shown a synthetic CMD for the age of 
76 Myr,  and turn-off mass $M_{\rm TO}=5M_{\odot}$.
 The termination magnitude of the 
main sequence band fits the
observational data.
However, the luminosity interval spanned by the blue loop does not agree
with the observed CMD. Theoretical luminosities are indeed brighter than
observed. Contrary-wise, if we impose that the mean luminosity of
red giants is matched, the age increases to $110$ Myr but 
the observational termination magnitude of the main sequence 
is about $1$ mag brighter than the theoretical one. In Fig.  
\ref{fig:classic_single_ILF} we present 
the comparison between the observational (dashed line) and the 
theoretical N-ILF 
(solid line) for the population of $76$ Myr in 
Fig. \ref{fig:classic_single_CM}. The stellar models in use yield a ratio 
$\Sigma N_{\rm MS}/N_{\rm PMS}$ lower than the observational  one.
Since $N_{\rm PMS}$ is taken equal to the number of red giants detected 
in NGC 1866, this implies that the classical semi-convective models 
produce more giants than observed (a long known  result).

\begin{figure}
\resizebox{\hsize}{!}{\includegraphics{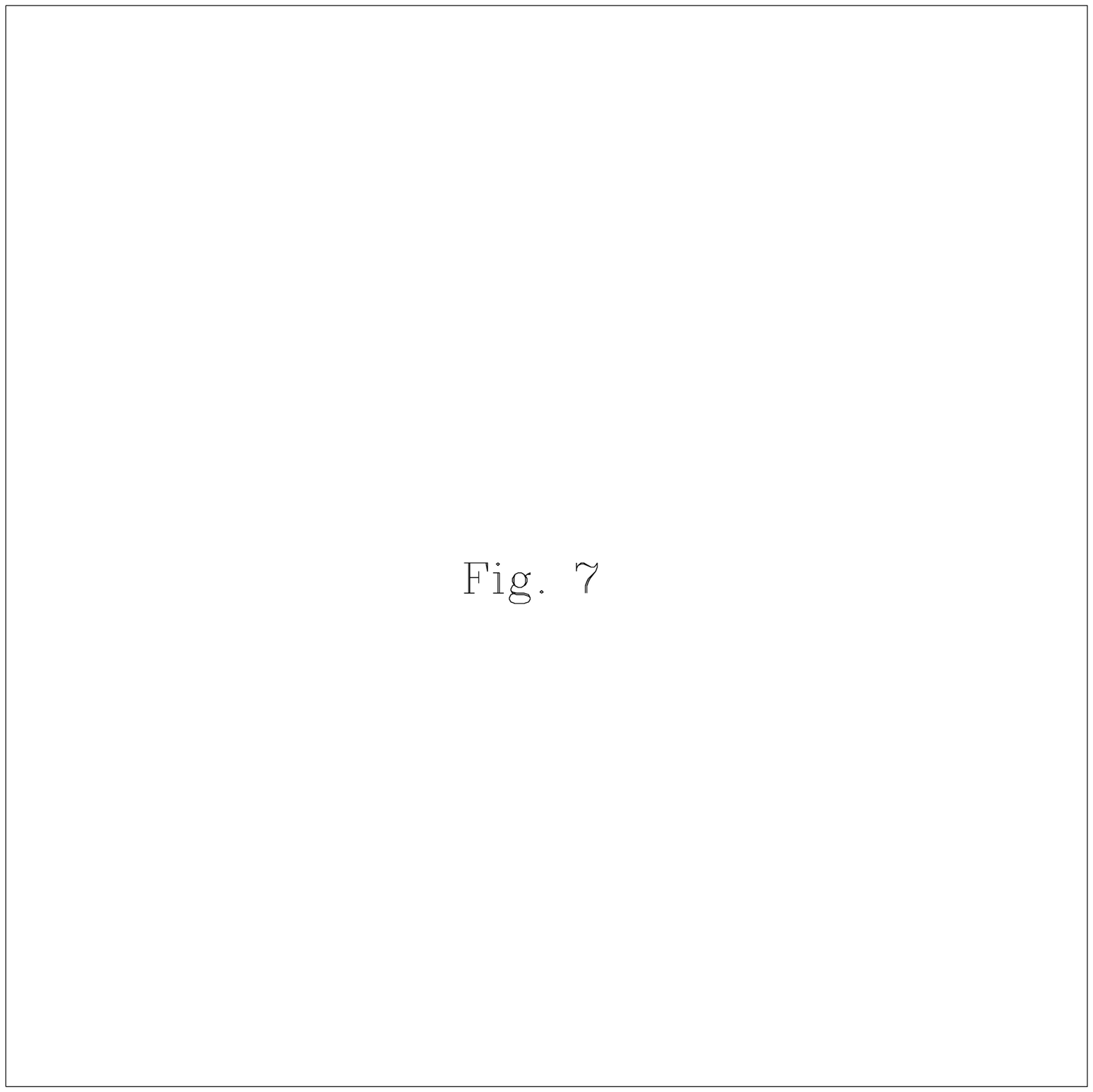}}
\caption{\small Synthetic CMD for the age of $160$ 
	Myr, obtained from models with overshooting and chemical 
	composition [$Y$=0.250,$Z$=0.008]}\label{fig:over_single_CM}
\end{figure}

\begin{figure}
\resizebox{\hsize}{!}{\includegraphics{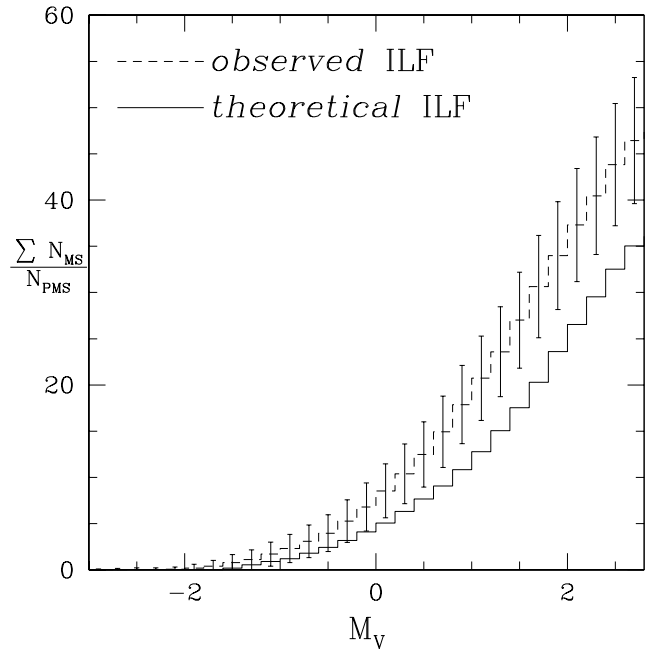}}
\caption{\small Observational (dashed line) N-ILF ---with the poissonian 
	error bars--- and theoretical N-ILF (thick line) for the 
	case of $160$ Myr shown in  Fig. 
	\protect{\ref{fig:over_single_CM}} 
	}\label{fig:over_single_ILF}
\end{figure}

\subsection{Models with overshooting}

As already anticipated commenting Fig. \ref{fig:iso_conf_cl_ov}, the 
longer 
lifetime and brighter luminosities of stellar models with overshooting 
impose  older isochrones to be adopted  in the CMD simulations. 
Figs. \ref{fig:over_single_CM} and \ref{fig:over_single_ILF} 
illustrate the synthetic CMD and the N-ILF, respectively,
and compare them  
with their observational counterparts for a population  $160$ Myr
old and   turn-off mass of $M_{\rm TO}=4.1M_{\odot}$.
Although a larger ratio
$\Sigma N_{\rm MS}/N_{\rm PMS}$ is 
obtained as compared with the classical case, still the agreement 
between theory and observations is not fully satisfactory:
indeed, while the region of the blue loop  is well fitted, 
the main sequence termination magnitude 
is about $1$ mag fainter than  observed. 

Alternatively, if a younger age ($ \simeq 100$ Myr) is adopted
to match  the main sequence termination luminosity,
theory and observations would disagree in the morphology of the blue 
loop. Furthermore, it would predict  too a high number 
 of main sequence stars ($\Sigma N_{\rm MS} \simeq 10000$) compared to 
 the observed number
of red giants ($N_{\rm PMS} = 100$), i.e. 
the ratio $\Sigma N_{\rm MS}/N_{\rm PMS}$ would be too high.

\subsection{Including  binary stars}

The effect of unresolved binary stars (UBS) on the CMD of stellar 
clusters has been investigated in a 
number of studies \citep[see][]{C89a,V91}.  

The main effect of UBS  on the main sequence morphology 
would be the appearance of a second sequence,
running parallel to the main sequence, but systematically 
cooler and with brighter luminosities ($\simeq 0.7$ mag for binary stars 
of equal mass). 
It goes without saying that in order to detect
this secondary sequence the number of UBS must be significant.

The percentage of UBS 
in open clusters of the Milky Way seems to amount to about 
30\,\% -- 50\,\% of 
the cluster population \citep{Mermi-Mayor89,Ca94}. Likely
the same percentage of UBS could exist also in young clusters
of the LMC. 
Strong observational evidence  of this comes 
from the recent study of \citet{El98} on NGC 1818, a young cluster  
similar to NGC 1866.  In the  CMD obtained with $HST$ data, the 
existence of a  ``double'' main sequence is soon evident.
\citet{El98} explain this 
feature by supposing the existence of  a population of UBS amounting to
 30--35\,\% of the total and with mass ratios in the range
$0.7$ to $1$. Systems with 
mass ratios different from these cannot be excluded. In any case they
would be hardly distinguishable from single objects.

Basing on the study by  \citet{El98}, we include in our 
simulations the same fraction 
of binary stars, to understand how they alter the above
conclusions both for  classical and  overshooting models. Unless
otherwise
specified, the mass ratio $q$ of the UBS included in the
simulations falls in the interval $0.7<q<1$ in agreement with 
\citet{El98}\footnote{Detailed simulations of the CMD of NGC 1818,
using the prescription for the population of UBS by \citet{El98} can be 
found
in \citet{Barmi01}. They are not shown here for the sake of brevity.}. 

The presence of UBS, shifting the main sequence 
termination  toward brighter luminosities, requires older 
ages. At the same time, the older 
the isochrone, the fainter is the blue loop. Therefore, under the action
of the two effects, the luminosity gap between the main sequence 
termination and  red giants is reduced and   the 
overall morphology of the CMD is better reproduced.

\begin{figure}
\resizebox{\hsize}{!}{\includegraphics{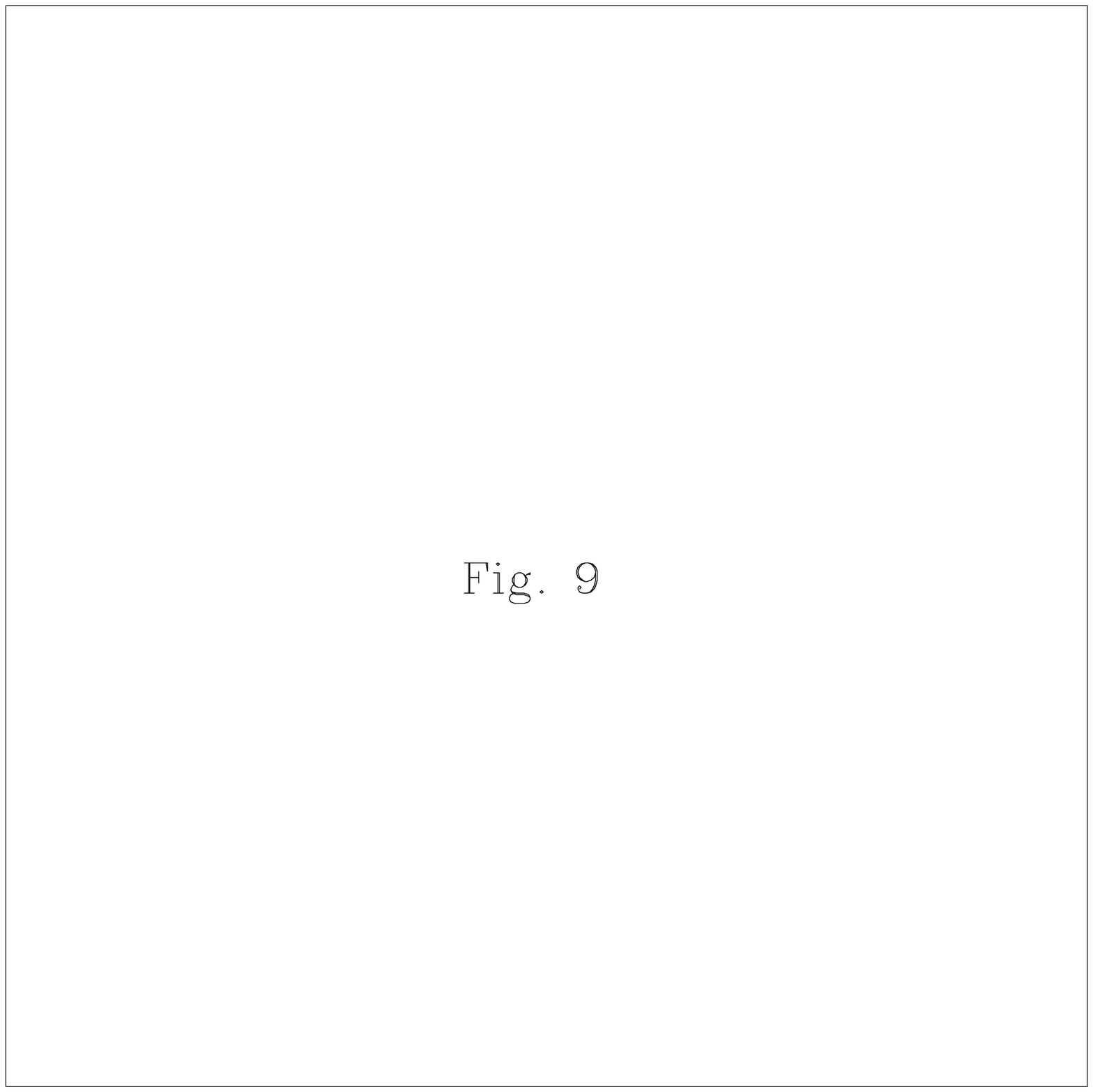}}
\caption{\small Synthetic CMD constructed with 
 classical models, chemical composition 
	[$Y$=0.250,$Z$=0.008], and presence of UBS. The age is 
	$105$ Myr; the percentage of UBS amounts to  $30\,\%$ 
}\label{fig:classic_binary_CM}
\end{figure}

\begin{figure}
\resizebox{\hsize}{!}{\includegraphics{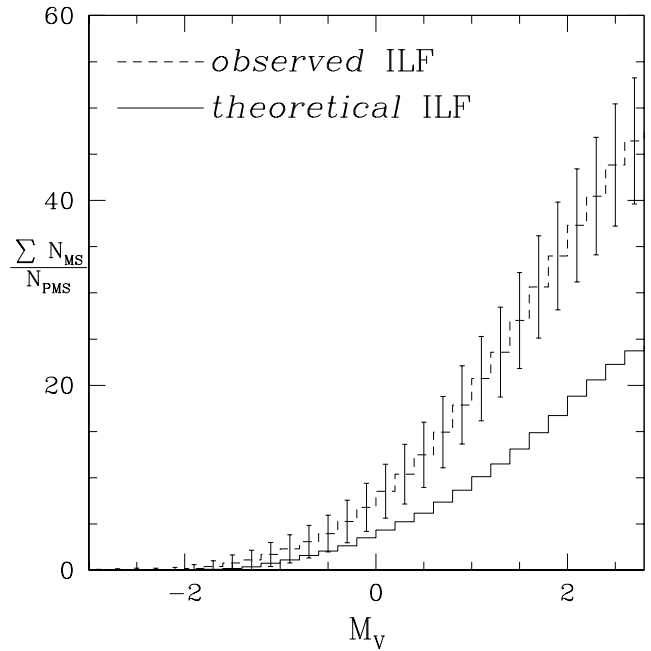}}
\caption{\small Observational N-ILF (dashed line) ---with the poissonian 
error 
	bars--- and theoretical ILF (thick line) for the
	case shown in Fig. \ref{fig:classic_binary_CM}:  classical 
	models, age 105 Myr, and  $30\,\%$ of  UBS.}
	\label{fig:classic_binary_ILF}
\end{figure}

\subsubsection{Binary stars in the classical picture}
In Fig. \ref{fig:classic_binary_CM} we present the synthetic CMD for the
age of $105$ Myr (turn-off mass $M_{\rm TO}=4.3$ M$_{\odot}$), 
in which $30\,\%$ of UBS are included. 
Compared to the results obtained simulating only single stars, 
the agreement with the observed CMD has much improved, 
but the theoretical N-ILF 
(the thick line in Fig. \ref{fig:classic_binary_ILF}), 
clearly suffers  the same difficulty already encountered neglecting the 
effects
of binary stars, i.e. too low a ratio $\Sigma N_{\rm MS}/N_{\rm PMS}$. 
The same simulation has been repeated adopting the same percentage 
of binaries ($30\,\%$) but supposing that the mass ratio is distributed
 according to 
a Gaussian curve, as suggested by  \citet{Te99}. 
No significant difference is noticed
both in the CMD morphology  and  N-ILF.

\subsubsection{Initial mass function}
\begin{figure}
\resizebox{\hsize}{!}{\includegraphics{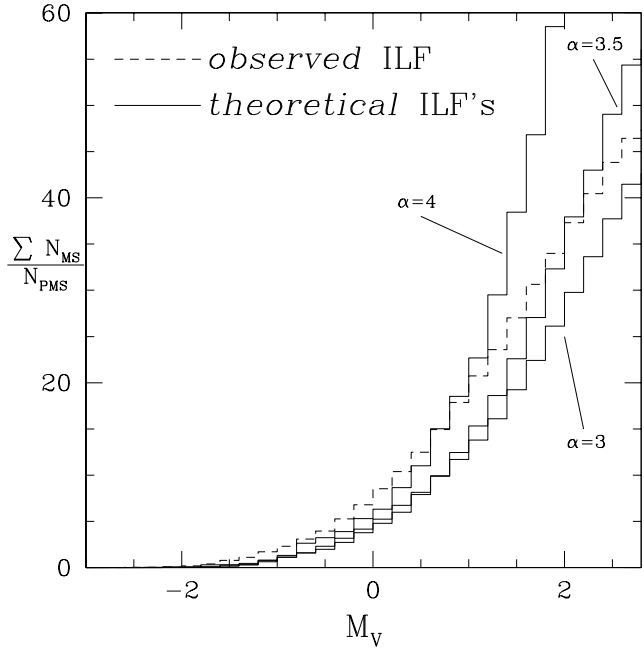}}
\caption{\small Comparison between the observational N-ILF (dashed line) 
and 
	the theoretical ones (thick lines) 
	obtained in the classical picture for a population of 
	$105$ Myr and  three different slopes for 
	the Salpeter IMF, namely $\alpha =3, \alpha =3.5, \alpha =4$. The 
	initial chemical composition, [$Y$=0.250,$Z$=0.008], the 
	number of post main sequence stars ($N_{\rm PMS}=100$), and 
	the fraction of binaries present in the sample ($30\,\%$) are 
	constant 
	in the three simulations}\label{fig:slopes_ILF}
\end{figure}

{\it How do the above results  depend on the particular slope 
($\alpha =2.35$) adopted for the Salpeter IMF?} 
To cast light on this topic we have repeated 
our simulations varying $\alpha$ and  keeping 
constant all other parameters.  The analysis shows that while 
the CMD morphology is scarcely affected, the N-ILF is sensitive to 
variations of $\alpha$.
As illustrated in Fig. \ref{fig:slopes_ILF}, 
the N-ILF gets steeper at increasing $\alpha$. 
We note that, in order to 
reach agreement between observational and theoretical N-ILFs using 
the classical semi-convective
models, one should adopt $\alpha$=(3.5 -- 4), This would  
favor  the formation of low mass stars  still located on the 
main sequence. The problem is whether or not such high values of 
$\alpha$ are acceptable. The present results    
 re-confirm what already pointed out in the earlier studies by 
\citet{BM83} and  \citet{C89a}.

\begin{figure}
\resizebox{\hsize}{!}{\includegraphics{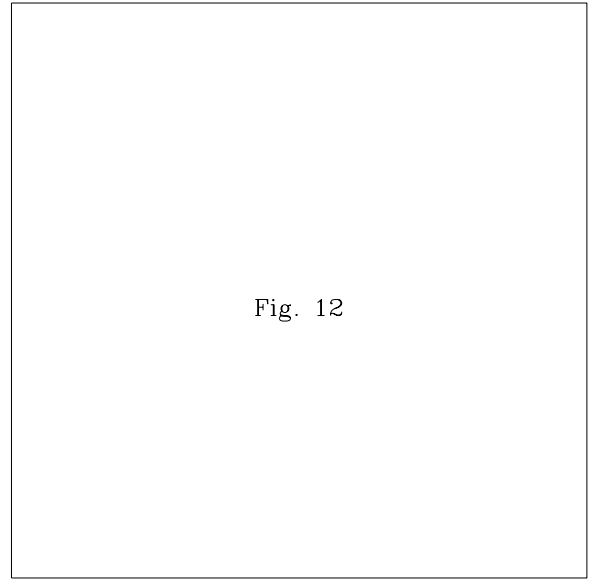}}
\caption{\small Synthetic CMD for  
	overshooting models:   age  of $148$ Myr 
	 $30\,\%$  of UBS,  and chemical composition 
	[$Y$=0.250,$Z$=0.008]}\label{fig:over_binary_CM}
\end{figure}

\begin{figure}
\resizebox{\hsize}{!}{\includegraphics{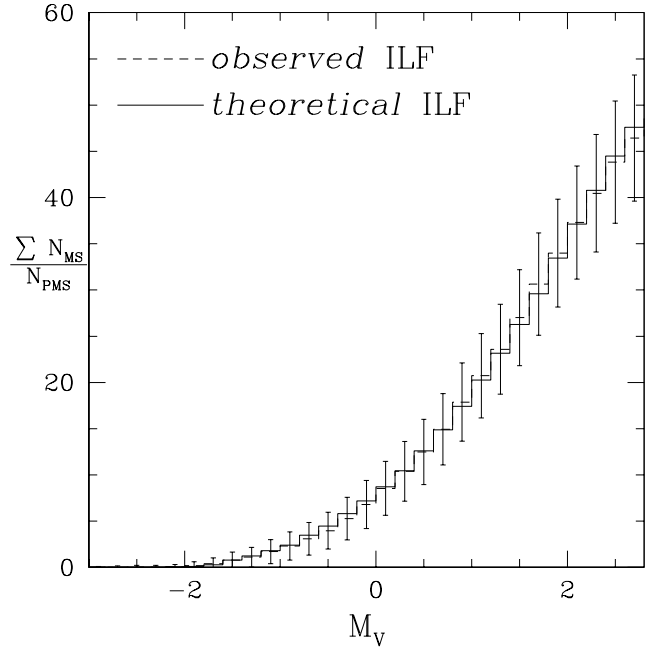}}
\caption{\small Observational N-ILF (dashed line) --with the Poissonian 
	error bars-- and theoretical N-ILF (thick line) for the $148$ Myr old
	population shown in Fig. \ref{fig:over_binary_CM}: overshooting 
	models
	and $30\,\%$ of UBS}\label{fig:over_binary_ILF}
\end{figure}

\subsubsection{Binary stars in the overshooting picture}

The synthetic CMD presented in Fig. \ref{fig:over_binary_CM} refers to 
stellar models with convective overshooting and simulates 
 a population  $148$ Myr old (turn-off mass 
$M_{\rm TO}=4.2$ M$_{\odot}$) 
containing  $30\,\%$ of UBS.
We note that the luminosity gap between the red giant region and the main 
sequence termination  has now disappeared. The CMD of NGC 1866 is 
perfectly reproduced.  The most important result is, however,  the N-ILF 
shown in  Fig. \ref{fig:over_binary_ILF}, which fully agrees with the 
observations. In other words, the ratio $\Sigma N_{\rm MS}/N_{\rm PMS}$ 
predicted by stellar models with convective overshoot together with
the presence of a certain fraction of binary stars fully agrees
with that obtained from star counts.

\begin{figure}
\begin{minipage}{0.5\textwidth}
\resizebox{\hsize}{!}{\includegraphics{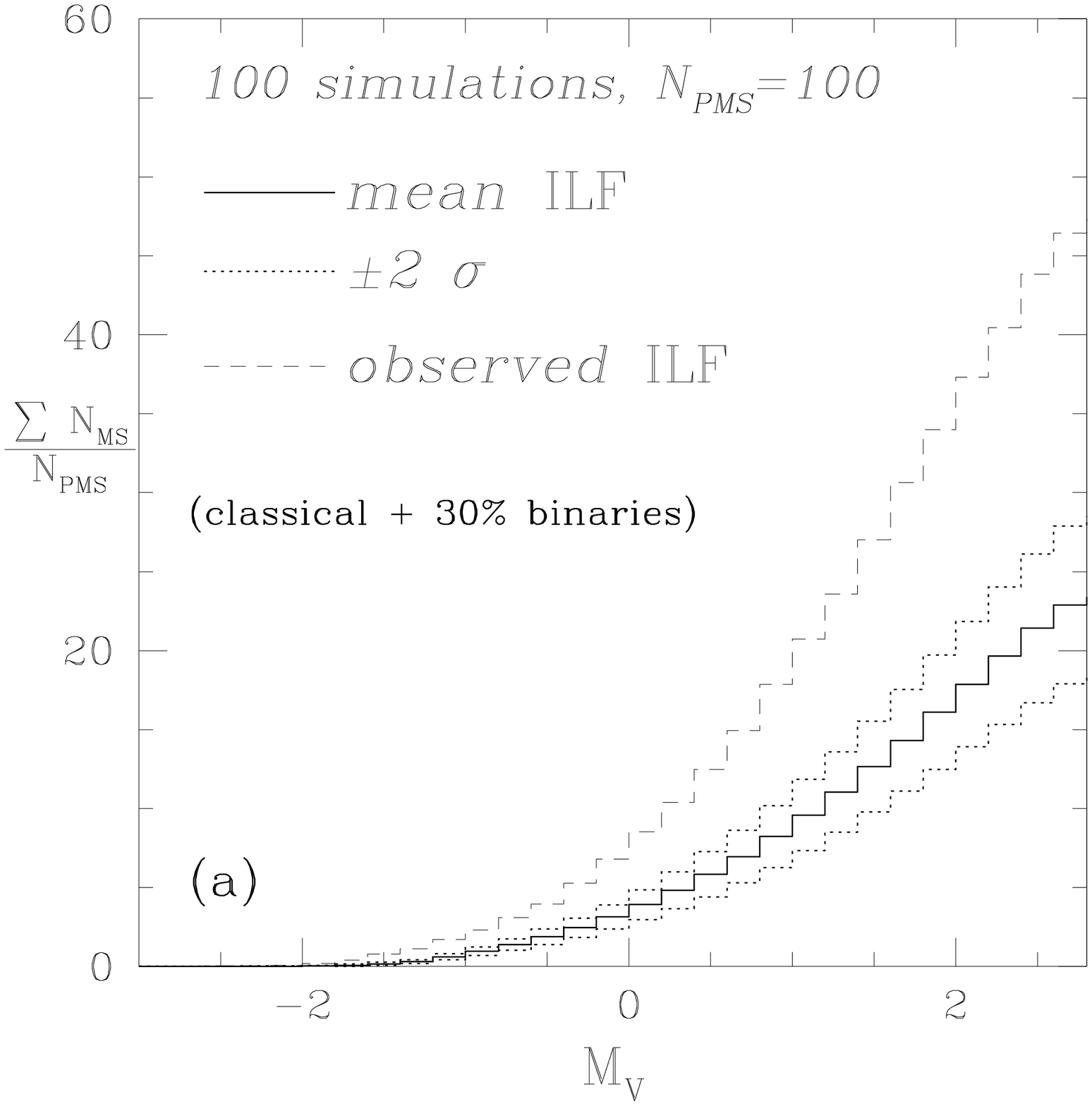}}
\end{minipage}
\begin{minipage}{0.5\textwidth}
\resizebox{\hsize}{!}{\includegraphics{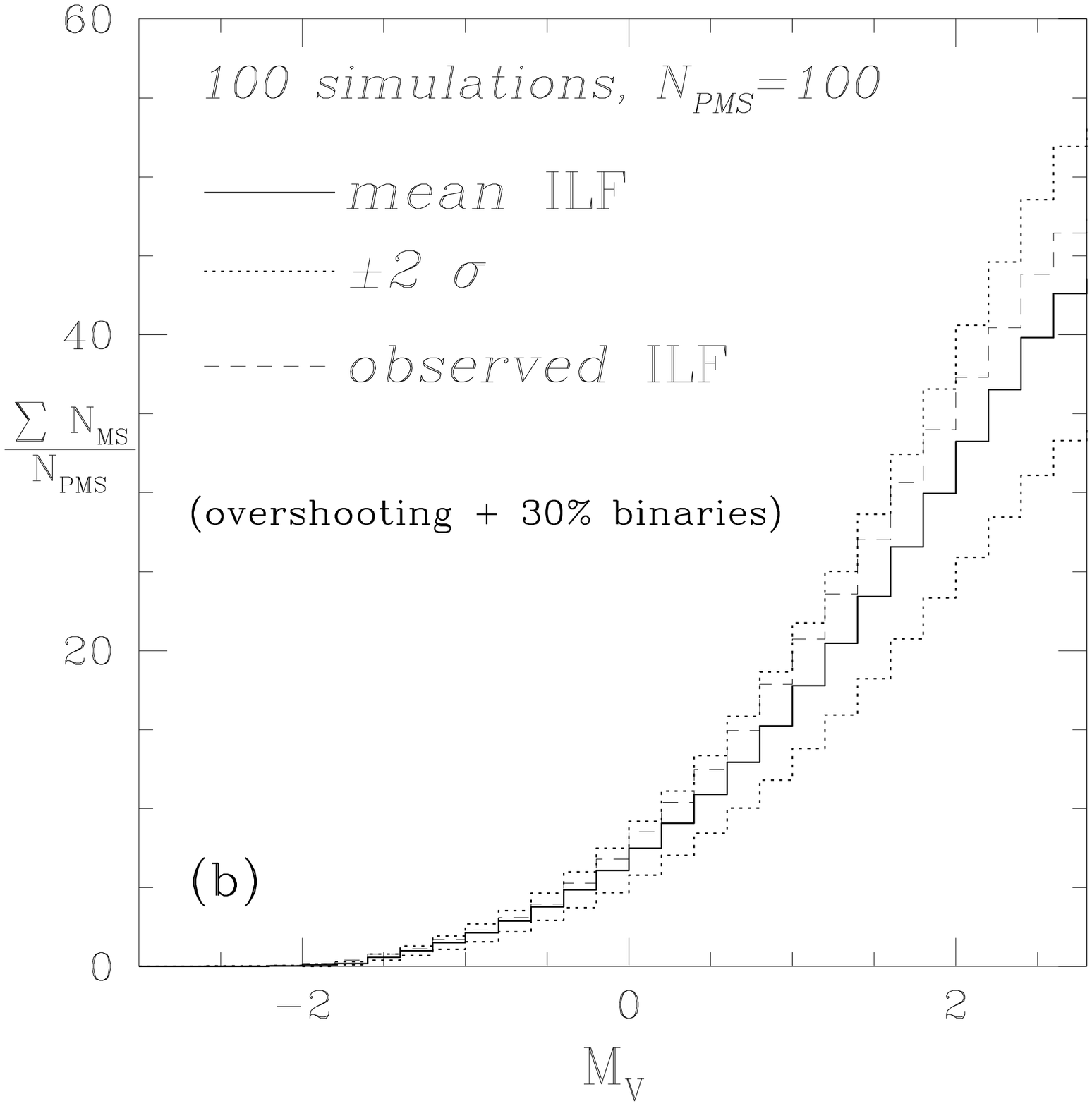}}
\end{minipage}
\caption{Mean N-ILF (thick line) and related $2 \sigma$ levels of 
	expectancy (dotted lines) resulting from 100 simulations of 
	the same population including  $30\,\%$  of UBS. 
	The chemical composition  is [$Y$=0.250,$Z$=0.008].
	With the classical scheme (top panel) and age of 105 Myr
	the observational N-ILF 
	(dashed line) is  steeper than the theoretical one; for the 
	overshooting scheme (bottom panel) and age of 148 Myr,
	the observational N-ILF 
	lies inside the $\pm 2 \sigma$ range around the theoretical 
	prediction
	}\label{fig:ILF_deviations}
\end{figure}

\subsection{Effects of stochastic fluctuations in the IMF}
The discussion of the ILF we have presented so far, refers to a 
particular
simulation of the CMD, which does not explicitly take into account 
possible
effects of stochastic nature caused by the finite number of stars in a 
cluster. It is worth recalling here that the concept itself of IMF
allows for important fluctuations to occur around the ideal distribution
of stars as a function of the mass.
The problem may be particularly severe for $N_{\rm PMS}$ which in most 
cases
is a small number. In other words the question to be addressed is whether 
stochastic effects may change  $N_{\rm PMS}$ with respect to 
$\Sigma N_{\rm MS}$ or
conversely, whether $N_{\rm PMS}$ which has been used as the key 
constraint to
our simulated CMDs and ILFs is generated by a unique population of main
sequence stars or if many others are equally eligible.

To cast light on this issue, we have performed many simulations of
our reference CMD keeping constant all parameters and letting the 
stochastic
nature of the IMF develop thanks to the Monte Carlo technique we have 
adopted. 
To this aim, 100 simulations of the same CMD are calculated in order to 
get the mean N-ILF and its standard deviation  $\sigma$. 
 
For the sake of brevity, we illustrate here the results only for 
the cases 
of classical models (Fig. \ref{fig:ILF_deviations}a)
 and overshooting models (Fig. \ref{fig:ILF_deviations}b),
in which $30\,\%$ of binary stars are included.
No variations to the 
previous conclusions seem to appear. With classical semi-convective
models the observed N-ILF (dashed line) 
runs above the  $\pm 2 \sigma$ interval (dotted lines) around the mean 
N-ILF (thick line). In contrast, with the overshooting models
theoretical and observational N-ILFs agree at the $\pm 2 \sigma$ level.
Considering the errors affecting the observational N-ILF, the 
agreement is extremely good.

\subsection{Age dispersion}
There is a final effect to be considered, i.e. the possibility that stars 
in 
the cluster are born over a finite time interval, i.e. that  a 
significant age spread exists.

\begin{figure}
\begin{minipage}{0.5\textwidth}
\resizebox{\hsize}{!}{\includegraphics{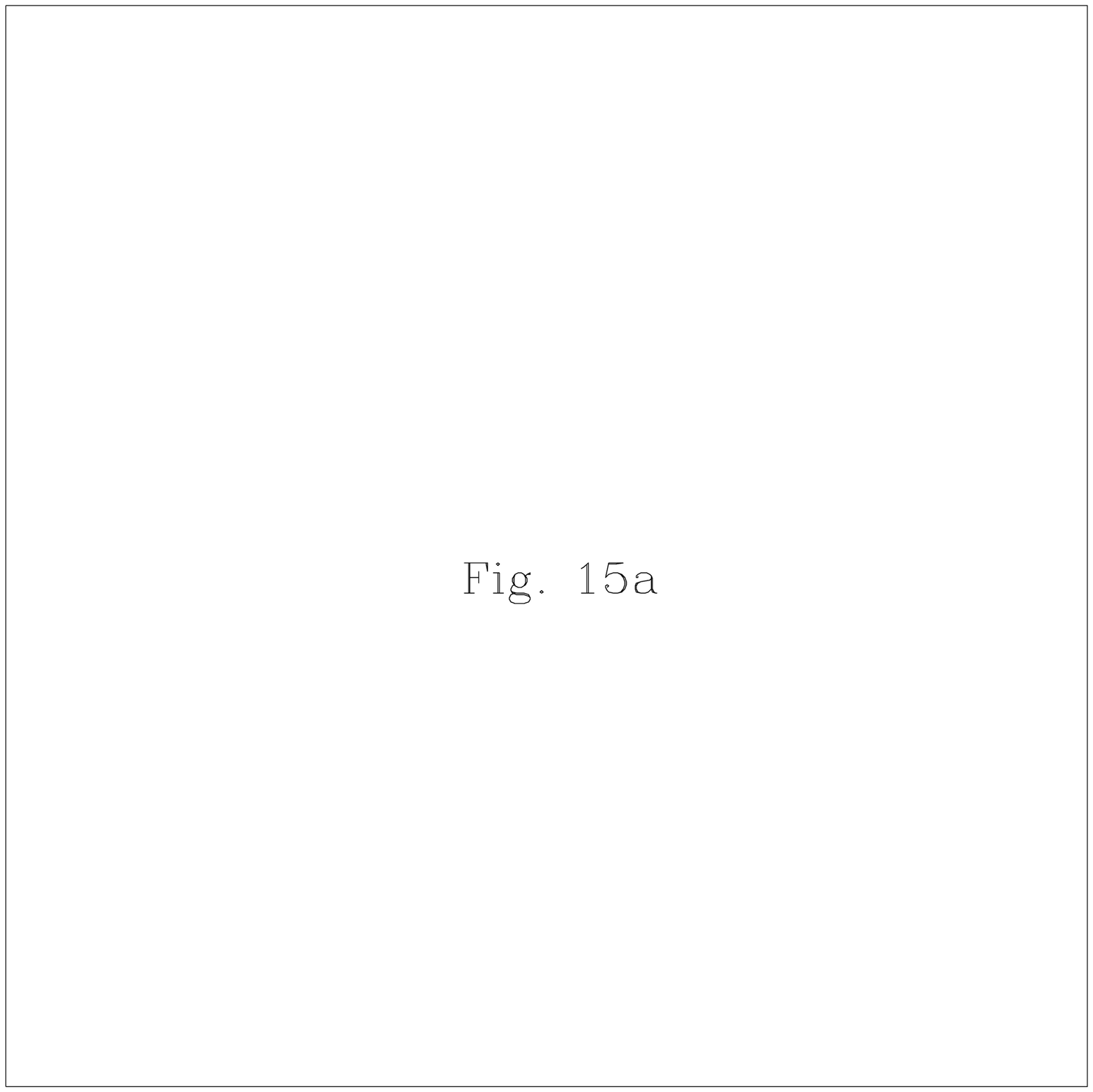}}
\end{minipage}
\begin{minipage}{0.5\textwidth}
\resizebox{\hsize}{!}{\includegraphics{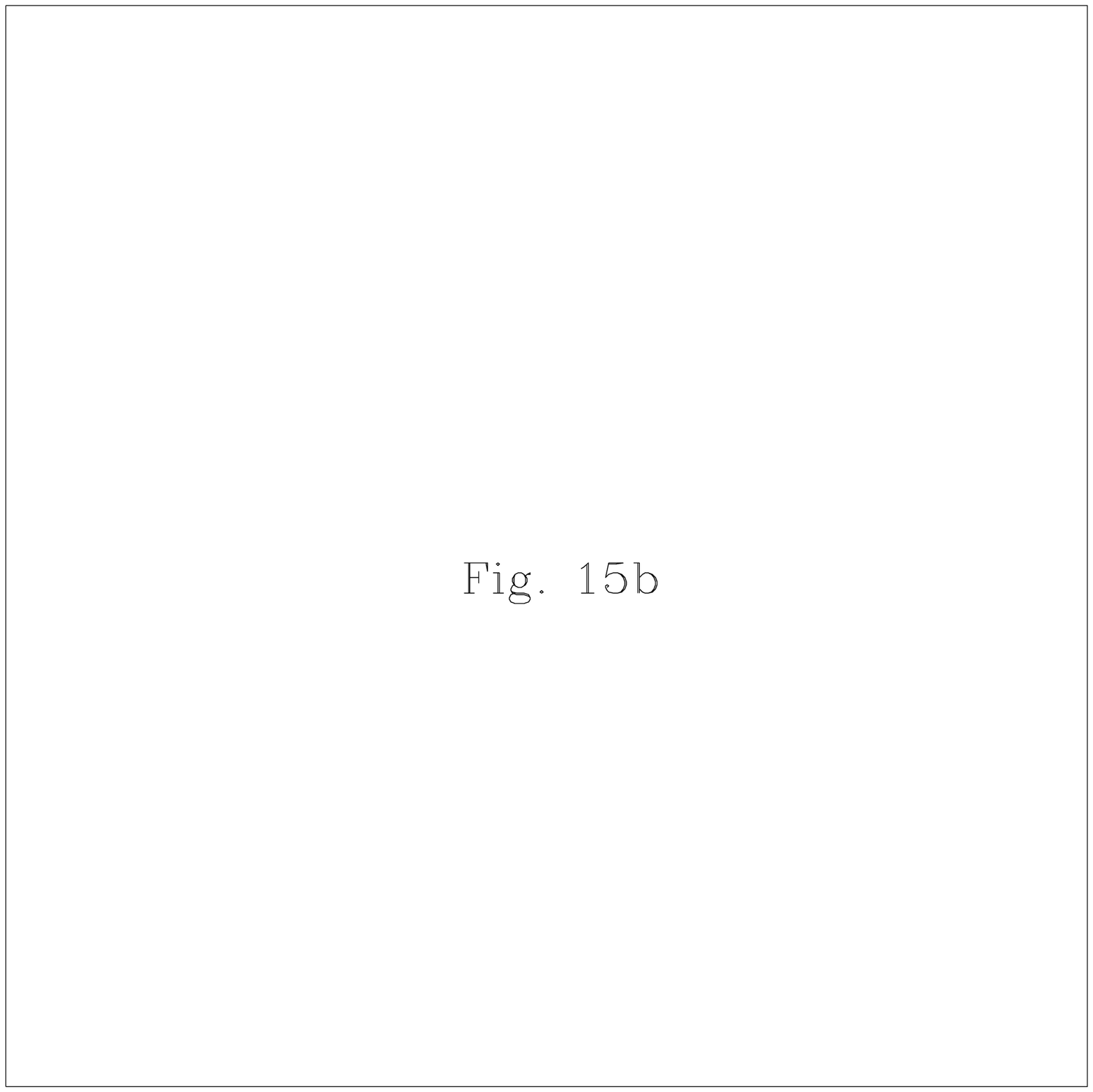}}
\end{minipage}
\caption{\small Comparison between the CMD 
	of NGC 1866 \citep{Te99} (top panel) and  our best-fit case 
	(bottom panel), 
	characterized by stellar models with convective overshooting,
	age in the  interval  140--160 Myr,  $30\,\%$ of UBS. 
	The initial chemical composition is [$Y$=0.250,$Z$=0.008]. The 
	intrinsic distance modulus  is $(m-M)_{\rm 0}=18.5$.}
	\label{fig:HR_age_spread}
\end{figure}

Starting from our best-fit CMD, obtained from overshooting models with 
age of 
$\simeq 148$ Myr and including $30\,\%$ of binary stars, adding an 
age spread of about $20$ Myr (about 10\,\%) 
would further improve the CMD, especially in
the region of red giant stars. This final CMD is presented in Fig.
\ref{fig:HR_age_spread}b, and compared with that of NGC 1866 
(Fig. \ref{fig:HR_age_spread}a). This small spread in the age
while giving the final make-up of the CMD does not affect the N-ILF,
and the picture outlined in Fig. \ref{fig:ILF_deviations}b still holds.

\subsection{Major uncertainties} 
Looking at CMD of Figs. \ref{fig:HR_age_spread}b and 
\ref{fig:HR_age_spread}a, we see that while overall 
agreement is reached especially as far as 
the magnitudes and colors  of the main 
sequence stars, and 
the mean luminosity of the stars in the blue loop are concerned, 
there is still 
a marginal disagreement in the colour range 
spanned by the red giant stars, i.e. the extension of the blue loop.
This, however, is 
known to depend on several factors: (i) the overshooting parameters 
$\Lambda_{\rm c}$ assumed for the nuclear region 
\citep[see][]{C89a} and 
the one adopted for the envelope, $\Lambda_{\rm e}$ \citep[see][]{Al91};
(ii) the rate of the nuclear reaction 
$^{12}{\rm C}(\alpha, \gamma)^{16}{\rm O}$ --which determines 
the abundances of $^{12}{\rm C}$ and $^{16}{\rm O}$ at the end of 
central 
He-burning phase-- the higher the rate, the bluer 
the loop \citep{Be85}. 
{\it It is worth noting, however, that these 
uncertainties on the extension of the blue loop do not affect the
star counts at the base of the ILF.}

On the observational side, apart from photometric errors, completeness 
correction, and  fraction of UBS that have already been taken into 
account, one has to 
check whether the results could depend in a crucial way on the particular 
calibration  adopted to convert luminosities and effective 
temperatures into magnitudes and 
colours of the $UBVRI$ system. 
To clarify this point, we have also used the empirical 
conversions by \citet{Alo99}, who present tabulations 
of colours and bolometric corrections as a function of the 
effective temperature 
and metallicity [Fe/H] of the stars. However, comparing  isochrones 
calculated with this transformation and those  obtained with the 
\citet{Be94} no appreciable difference is noticed.

Some uncertainty could be related to  the particular choice for the
percentage of UBS ($30\,\%$) and  their mass ratios  $0.7 < q < 1$ we
have adopted. {\it How could a change in 
these parameters affect our conclusions? } To cast light on the 
issue
we have performed additional simulations with different percentages and 
mass ratios. Increasing the 
binary fraction up to about $40\,\%$, and extending 
the mass ratio interval down to $0.5 - 0.6$, the CMD and N-ILF are still 
in  agreement with the cluster data. However, assuming an even larger 
percentage of
binary stars ($>40 - 45 \,\%$), the CMD  gets worse as 
the  main sequence band spreads too much toward red colors.

\section{Why do we disagree with \citet{Te99}? } \label{why:testa}
\label{sec:why}
  
The results of our analysis wholly disagree with those by 
 \citet{Te99}, despite the fact that we are using the same observational 
 data
for  NGC 1866. Their main conclusion is indeed that classical stellar 
models
without overshooting fit the CMD and ILF of NGC 1866 at the age of 100 
Myr
provided that a population of binary stars amounting to about   
$30\,\%$ of the total is included. In contrast our main conclusion is 
that even 
including the same percentage of binary stars only models with
overshooting simultaneously reproduce the CMD, the ILF, and the correct
ratio $\Sigma N_{\rm MS}/N_{\rm PMS}$. The age now is about 150 Myr.

{\it What is the cause of the opposite conclusions?} There are several 
factors concurring to the final result:

(i) The different procedure applied to correct star counts for 
photometric
completeness. The ILF of \citet{Te99} is over-estimated due to the
the fact that instead of summing up the DLF corrected for 
completeness in every magnitude bin, they apply the correction procedure
to every step of the ILF.

(ii) The different normalization procedures of the simulations.
In the present study, the normalization parameter of 
the simulated CMD and ILF is 
the total (observed) number of red giants $N_{\rm PMS}$. It is worth 
recalling
that the N-ILF is given by $\Sigma N_{\rm MS}/N_{\rm PMS}$. The 
advantage of this type of normalization has already been explained by
 \citet{C89a} and briefly summarized in previous sections.
In contrast \citet{Te99} assume that each simulation is complete 
as soon as a given total number of stars with magnitude brighter than 
$M_V=2.6$ is reached. They justify this choice 
by saying that their main sequence is well defined and complete
up to three magnitudes fainter than the turn-off, and assume the total 
number of stars observed in this magnitude interval as the 
{\it normalization parameter}
of the simulations instead of the giant (post main sequence) stars.

(iii) This choice of the normalization factor is, however, not physically 
sound because a deeply observed, well populated main sequence does not tell
much about the inner structure, evolutionary rates, lifetimes etc. of its 
stars.
Indeed the distribution of the stars along the main sequence  almost
exclusively depends on the IMF.
Therefore, the choice of a ``normalization parameter'' 
 insensitive to the size of the convective core simply precludes all 
 chances
of discriminating between the two evolutionary schemes under examination.
Simulations of CMDs and ILFs based on this criterion may fail to 
reproduce the
correct ratio $\Sigma N_{\rm MS}/N_{\rm PMS}$.

(iv) In relation to this, is also the fact that the total number of
stars brighter than $M_V=2.6$  on which the simulations are normalized
is determined by the procedure of 
 correcting for photometric incompleteness. We have already argued that the
\citet{Te99} method is inconsistent, and overestimates the total 
number of stars in the main sequence. Since their simulated
 CMDs and ILF do actually depend on this parameter, there is no way 
of becoming aware of the internal inconsistency.

Considering all the differences between the present and \citet{Te99}
analyses, we should recognize that {\em 
the contrasting results are mainly due to the
different way of normalizing the ILF: we choose a normalization factor
(number of giants) that is sensitive to overshooting, 
whereas Testa et al.'s (1999) choice (stars brighter than $M_V=2.6$) is 
essentially insensitive to this effect}.
In fact, no clear discrimination between classical and overshooting 
models is expected with the latter method.

(v) 
To better understand this point of controversy
we have repeated our simulations strictly following the procedure 
described
by \citet{Te99}: each one contains the same total number of stars 
brighter than $V=22$ as the observational data, i.e.\ 4879. 
The results are displayed in Fig.~\ref{fig:tuttotesta}, 
limited to the few cases we have already examined. The layout of 
Fig.~\ref{fig:tuttotesta} is organized in vertical and horizontal groups 
 of panels. Panels in the same vertical row correspond to the 
case as indicated by the top heading of the row. Panels in horizontal 
rows
show different types of LFs as indicated: the DLF in the bottom, the 
total
ILF (main sequence and giant stars) in the middle panels, and the  ILF 
for the sole main sequence stars in the top panels.
The results can be commented as follows:

\begin{figure*}
\begin{center}
\begin{minipage}{0.80\textwidth}
\resizebox{\hsize}{!}{\includegraphics{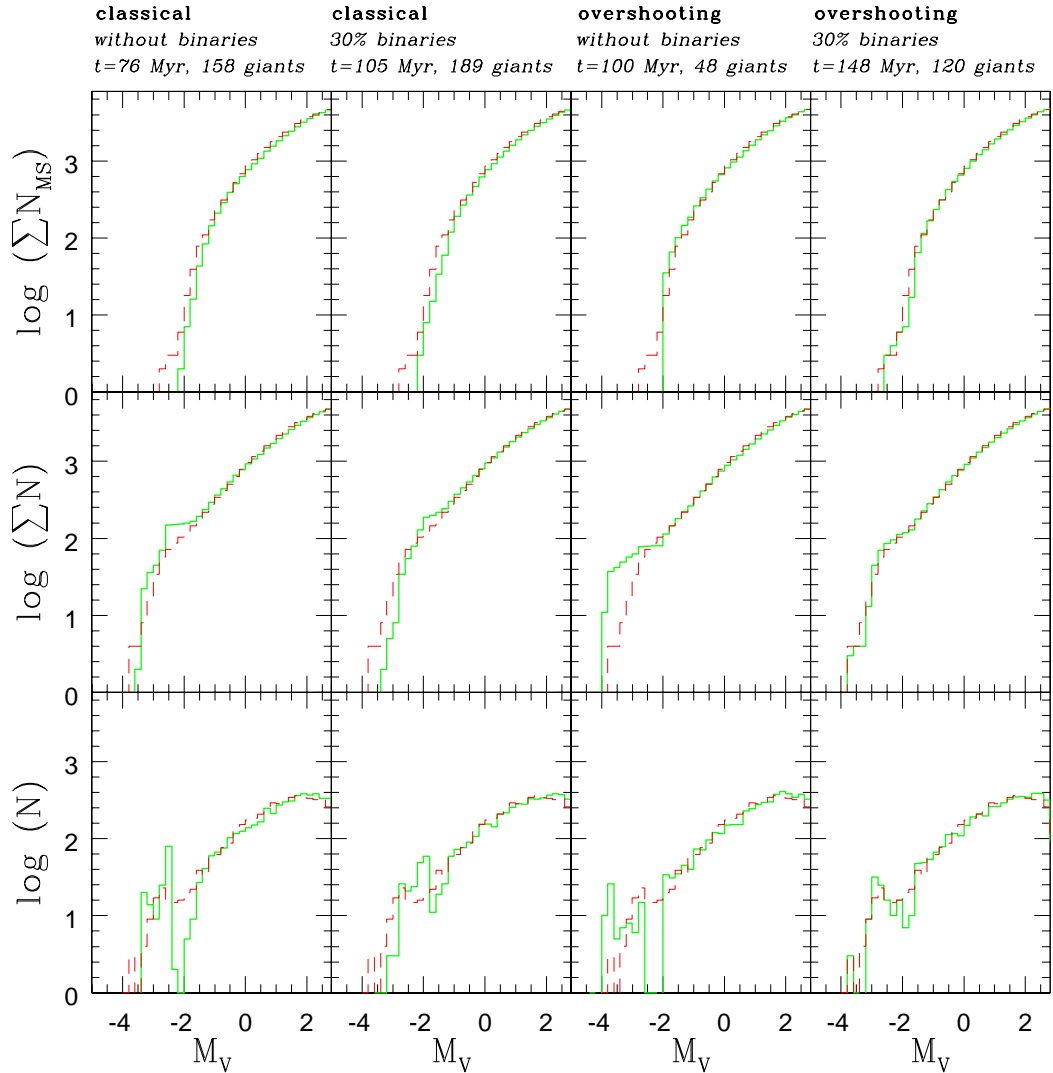}}
\end{minipage}
\end{center}
\caption{\small Summary of the four cases under consideration: 
the heading of each column lists the type of stellar model
(classical or overshooting), the percentage of unresolved binaries, 
the age, and the number of giant stars.
The three horizontal rows display the LFs as shown in 
\citet{Te99}: 
the top row is the integrated LF for the sole main sequence stars in the
simulation, the middle 
row is the integrated LF  for all stars in the simulation, finally the
bottom row is the DLF for the whole sample.
 In each panel the solid line is the theoretical LF or DLF, 
whereas the dashed line is the same but for the
observational data.
Each simulation contains a {\it total number of stars } brighter than  
 $V=22$ equal to $4879$}
	\label{fig:tuttotesta}
\end{figure*}

\begin{itemize}

\item The DLFs in the bottom horizontal row make evident the 
luminosity gap between the main sequence termination and 
the giant stars, located at $M_{V}\simeq-2$.
On the observational side, the gap is almost invisible, whereas 
on the theoretical side, the inclusion of binaries wipes out the gap
in both classical and overshooting models. 
All models seem to reproduce the main sequence part of the DLF 
($M_{V}\ga-1$) equally well.

\item The total ILFs, i.e. inclusive of the post main sequence stars,
are shown in the central horizontal row.
For the models without binaries, the luminosity interval between the main 
sequence termination and the giant stars appears as a plateau in the
ILF at $M_{V}\simeq-2$. To the left of this plateau, the detailed
shape of the ILF is determined mainly by the location of 
giant stars: a slight over luminosity of the giants (with respect to 
observations) causes an apparent excess of stars at $M_{V}<-2$, 
which can be noticed in the cases of classical models
with and without binaries, and overshooting models without binaries.
In the case of overshooting models with binaries, giant stars are
predicted at the right magnitude level, and no such discrepancy can be
noticed in the ILF.

But it is important to notice that the ``excess'' suggested by the 
ILF does not necessarily correspond to a real excess of evolved stars. 
This is evident if we look at 
the total number of giants in each simulation (see the labels at the top 
of the figure): in the case of classical models there is a real excess 
of giants (158 and 189 depending on whether or not binaries are 
included), which has no observational counterpart.
The situation is the opposite for models with convective overshoot 
and no binary stars. Now the simulated number of giants is 48 (about
half of what observed).
Finally, models with overshooting and binaries predict the correct 
number of post main sequence stars (120, to be compared with the
100 observed).

\item The ILF of main sequence stars  (top horizontal row) does not 
show substantial differences passing from one case to another.
This simply reflects the constancy of the IMF. There are small 
differences at the turn-off level, however too small to be significant.

\end{itemize}

>From the plots in Fig.~\ref{fig:tuttotesta}, there is a hint that models 
with overshooting and binary stars are in slightly better position, 
because they fit well the observed LFs in all panels. 
However, the most indicative result is not derived from the
shape of the different LFs, but comes from the expected number
of giants in the simulations: models with overshooting and binary stars 
are {\em the only ones} to predict the observed number of giant stars.

In other words, with the kind of plot used by \citet{Te99},
one cannot pin down the correct value of the ratio 
$\Sigma N_{\rm MS}/N_{\rm PMS}$. Finally, we like to call attention
to the fact  
that {\it even with the \citet{Te99} method,
no strong indication is found that classical stellar models are to be
preferred.}

\section{Conclusions}\label{sec:conclusions} 
The main goal of this study is to compare with observational data
the predictions for CMD and ILF obtained from two different types of 
stellar models: the classical ones in which  the extension of convective 
zones is defined by the 
Schwarzschild criterion (whenever required semi-convection is also taken
into account), and the models with overshooting (calculated 
accordingly to the \citet{Br81} formalism), which allow the 
penetration 
of convective elements into the surrounding formally stable layers. 
The ultimate goal is to cast light on which type of stellar models find
better correspondence with the physical structure of real stars.

The main result that  bears very much on the 
comparison with the observational data
is the net decrease 
--about $55 - 60\,\%$-- of 
the lifetime ratio $t_{\rm He}/t_{\rm H}$ 
passing from classical to overshooting models.

The analysis begins 
with correcting in the proper way the star counts and the LFs 
to get the LF of main sequence stars normalized 
to the number of evolved stars (N-ILF). The N-ILF 
yields the ratio 
$\Sigma N_{\rm MS}/N_{\rm PMS}$ of main sequence to giant stars, and  
directly measures  the ratio  $t_{\rm He}/t_{\rm H}$ \citep[see][]{C89a}.

Many simulations of the  CMD and LF  of NGC 1866  have 
been performed,  in which not only the effect of the different stellar 
models but also those given by the presence of binary stars is taken 
into account. The results can be summarized as follows:

\begin{itemize}

\item{
The observed ILF cannot be matched at all by the classical models, 
showing too large a ratio of evolved to main sequence stars.
The reason for the discrepancy between our results and the conclusion 
reached by 
\citet{Te99} has already been discussed and will not be repeated here.}

\item{The situation is the opposite for the case of models with 
overshooting.
It is slightly worse neglecting unresolved 
binaries and fully satisfactory with binaries. }

\item{The presence of unresolved binaries in the simulations both for
classical and overshooting models, allows to simultaneously match 
the luminosity of the main sequence termination and giant stars. }

\item{Binary stars act on the main sequence morphology in a way roughly 
mimicking the effect of an enlarged H-burning core, i.e. shifting the 
turn-off toward brighter luminosities and forcing us to use older 
isochrones  to fit the cluster data. However, they do not alter 
the ratio  $t_{\rm He}/t_{\rm H}$. }

\item{If all this can be taken as observational evidence for 
 more extended mixing in the cores of 
hydrogen burning stars, the stellar models with convective
overshooting constitute the simplest solution. They are indeed able 
to simultaneously match the CMD 
and ILF of NGC 1866, thus re-confirming what already 
found  in previous studies 
\citep{C89a,C89b,L91,V91,VaC92}.}

\item{
By adopting the  intrinsic distance modulus $(m-M)_{\rm 0}$=$18.5$ and the 
initial 
chemical composition [$Y$=0.250,$Z$=0.008], our best-fit of the CMD and 
ILF
of NGC~1866 is for an age in the range 140 to 160 Myr, turn-off mass  
$M_{\rm TO}$=$4.2$ M$_{\odot}$, 30 to  40\,\% of binary stars 
with mass ratio $q$ lying in 
the interval $0.6<q<1$. } 

\item{
Finally, it is expected that adopting slightly different values
for the distance modulus will not change the above conclusions.
Looking at the turn-off luminosity 
and its relation with the age, we estimate that  
$\Delta (m-M)_{\rm 0}$=$\pm0.1$ with respect to the adopted value
$(m-M)_{\rm 0}$=$18.5$ would change the age (and age range) by
$\Delta t \simeq \pm 10$
Myr. Small variations in the percentage of
binaries and/or age dispersion would immediately follow. }

\end{itemize}

New observational data 
will certainly improve the exact values of age, metallicity, binary 
star 
percentage so that the present values ought to be considered as 
provisional
estimates. Although  the new generation telescopes could 
make it possible to reach the 
innermost  regions of the cluster, scarcely or not at all affected by 
field star contamination, thus  providing us with better CMDs and ILFs,
 it is  unlikely that the present conclusions will change significantly.
{\it Real stars should indeed  possess bigger convective cores 
than predicted by the 
classical theory of stellar evolution.}

\vspace{0.5truecm}
\rightline{ {\it The civil war strikes again ...} }

\vspace{0.5truecm}
{\it Acknowledgments.}
We like to thank V. Testa and O. Straniero for the helpful clarifications 
about the observed luminosity functions of NGC 1866. 
This study has been financed by the Italian 
Ministry of Education, University, and Research (MIUR).


\begin{thebibliography}
\expandafter\ifx\csname natexlab\endcsname\relax\def\natexlab#1{#1}\fi

\bibitem[{{Alongi} {et~al.}(1991){Alongi}, {Bertelli}, {Bressan}, \&
  {Chiosi}}]{Al91}
{Alongi}, M., {Bertelli}, G., {Bressan}, A., \& {Chiosi}, C. 1991, A\&A, {\bf
  244}, 95

\bibitem[{{Alongi} {et~al.}(1993){Alongi}, {Bertelli}, {Bressan}, {Chiosi},
  {Fagotto}, {Greggio}, \& {Nasi}}]{Al93}
{Alongi}, M., {Bertelli}, G., {Bressan}, A., {Chiosi}, C., {Fagotto}, F.,
  {Greggio}, L., \& {Nasi}, E. 1993, A\&AS, {\bf 97}, 851

\bibitem[{{Alonso} {et~al.}(1999){Alonso}, {Arribas}, \&
  {Martinez-Roger}}]{Alo99}
{Alonso}, A., {Arribas}, S., \& {Martinez-Roger}, C. 1999, A\&AS, {\bf 140},
  261

\bibitem[{{Barmina}(2001)}]{Barmi01}
{Barmina}, R. 2001, in Thesis for the Master Degree in Astronomy, University 
of
  Padova, Italy

\bibitem[{{Becker} \& {Mathews}(1983)}]{BM83}
{Becker}, S. \& {Mathews}, J. 1983, AJ, {\bf 270}, 155

\bibitem[{{Bertelli} {et~al.}(1985){Bertelli}, {Bressan}, \& {Chiosi}}]{Be85}
{Bertelli}, G., {Bressan}, A., \& {Chiosi}, C. 1985, A\&A, {\bf 150}, 33

\bibitem[{{Bertelli} {et~al.}(1986){Bertelli}, {Bressan}, {Chiosi}, \&
  {Angerer}}]{Be86}
{Bertelli}, G., {Bressan}, A., {Chiosi}, C., \& {Angerer}, K. 1986, A\&AS, {\
bf
  66}, 191

\bibitem[{{Bertelli} {et~al.}(1994){Bertelli}, {Bressan}, {Chiosi}, {Fagotto}
,
  \& {Nasi}}]{Be94}
{Bertelli}, G., {Bressan}, A., {Chiosi}, C., {Fagotto}, F., \& {Nasi}, E. 199
4,
  A\&AS, {\bf 106}, 275

\bibitem[{{Bertelli} {et~al.}(1993){Bertelli}, {Bressan}, {Chiosi}, {Mateo}, 
\&
  {Wood}}]{Bw93}
{Bertelli}, G., {Bressan}, A., {Chiosi}, C., {Mateo}, M., \& {Wood}, P. 1993,
  ApJ, {\bf 412}, 160

\bibitem[{{B\"ohm-Vitense}(1958)}]{Bohm55}
{B\"ohm-Vitense}, E. 1958, Z. Astroph., {\bf 46}, 108

\bibitem[{{Bressan} {et~al.}(1981){Bressan}, {Bertelli}, \& {Chiosi}}]{Br81}
{Bressan}, A., {Bertelli}, G., \& {Chiosi}, C. 1981, A\&A, {\bf 102}, 25

\bibitem[{{Bressan} {et~al.}(1993){Bressan}, {Fagotto}, {Bertelli}, \&
  {Chiosi}}]{Br93a}
{Bressan}, A., {Fagotto}, F., {Bertelli}, G., \& {Chiosi}, C. 1993, A\&AS, {\
bf
  100}, 647

\bibitem[{{Brocato} {et~al.}(1989){Brocato}, {Buonanno}, {Castellani}, \&
  {Walker}}]{Broca89}
{Brocato}, E., {Buonanno}, R., {Castellani}, V., \& {Walker}, A. 1989, ApJS,
  {\bf 71}, 25)

\bibitem[{{Caldwell} \& {Coulson}(1985)}]{Cald85}
{Caldwell}, J. \& {Coulson}, I. 1985, MNRAS, {\bf 212}, 879

\bibitem[{{Canuto} \& {Mazzitelli}(1991)}]{Canuto91}
{Canuto}, S. \& {Mazzitelli}, I. 1991, ApJ, {\bf 370}, 295

\bibitem[{{Canuto} {et~al.}(1996){Canuto}, {Goldman}, \&
  {Mazzitelli}}]{Canuto96}
{Canuto}, V.~M., {Goldman}, I., \& {Mazzitelli}, I. 1996, ApJ, {\bf 473}, 550

\bibitem[{{Carraro} {et~al.}(1994){Carraro}, {Chiosi}, {Bressan}, \&
  {Bertelli}}]{Ca94}
{Carraro}, G., {Chiosi}, C., {Bressan}, A., \& {Bertelli}, G. 1994, A\&AS, {\
bf
  103}, 375

\bibitem[{{Chiosi}(1999)}]{Chiosi99}
{Chiosi}, C. 1999, in Stellar Structure: Theory and Test of Convective Energy
  Transport, eds. Alvaro Gimenez, Edward F. Guinan, and Benjamin Montesinos,
  ASP Conference Series, 173, ASP (San Francisco), 9

\bibitem[{{Chiosi} {et~al.}(1992){Chiosi}, {Bertelli}, \& {Bressan}}]{C92}
{Chiosi}, C., {Bertelli}, G., \& {Bressan}, A. 1992, ARA\&A, {\bf 30}, 235

\bibitem[{{Chiosi} {et~al.}(1989{\natexlab{a}}){Chiosi}, {Bertelli}, {Meylan}
,
  \& {Ortolani}}]{C89a}
{Chiosi}, C., {Bertelli}, G., {Meylan}, G., \& {Ortolani}, S.
  1989{\natexlab{a}}, A\&A, {\bf 219}, 167

\bibitem[{{Chiosi} {et~al.}(1989{\natexlab{b}}){Chiosi}, {Bertelli}, {Meylan}
,
  \& {Ortolani}}]{C89b}
---. 1989{\natexlab{b}}, A\&AS, {\bf 78}, 89

\bibitem[{{Cloutman} \& {Whitaker}(1980)}]{Cloutman80}
{Cloutman}, L. \& {Whitaker}, R.~W. 1980, ApJ, {\bf 237}, 900

\bibitem[{{Deardorff} {et~al.}(1969){Deardorff}, {Willis}, \&
  {Lilly}}]{Deardorff69}
{Deardorff}, J., {Willis}, G., \& {Lilly}, D. 1969, Fluid Mech., {\bf 35}, 7

\bibitem[{{Dominguez} {et~al.}(1999){Dominguez}, {Chieffi}, {Limongi}, \&
  {Straniero}}]{Dominguez99}
{Dominguez}, I., {Chieffi}, A., {Limongi}, M., \& {Straniero}, O. 1999, ApJ,
  {\bf 524}, 226

\bibitem[{{Elson} {et~al.}(1998){Elson}, {Sigurdsson}, {Davies}, {Hurley}, \&
  {Gilmore}}]{El98}
{Elson}, A., {Sigurdsson}, S., {Davies}, M., {Hurley}, J., \& {Gilmore}, G.
  1998, MNRAS, {\bf 300}, 857

\bibitem[{{Feast}(1989)}]{Feast89}
{Feast}, M. 1989, in The World of Galaxies, (ed.) {Corwin}, H.G. and and
  {Bottinelli}, L., (NY: Springer), 118

\bibitem[{{Feast}(2000)}]{Feast00}
{Feast}, M. 2000, astro-ph/0010590

\bibitem[{{Freytag} {et~al.}(1996){Freytag}, {Ludwig}, \&
  {Steffen}}]{Freytag96}
{Freytag}, B., {Ludwig}, H., \& {Steffen}, M. 1996, A\&A, {\bf 313}, 497

\bibitem[{{Girardi} {et~al.}(2000){Girardi}, {Bressan}, {Bertelli}, \&
  {Chiosi}}]{Gi00}
{Girardi}, L., {Bressan}, A., {Bertelli}, G., \& {Chiosi}, C. 2000, A\&AS, {\
bf
  141}, 371

\bibitem[{{Keller} {et~al.}(2000){Keller}, {Da Costa}, \& {Bessel}}]{Keller00
}
{Keller}, S., {Da Costa}, G., \& {Bessel}, M. 2000, astro-ph/0011285, to appe
ar
  in Astronomical Journal

\bibitem[{{Kurucz}(1992)}]{Ku92}
{Kurucz}, R.~L. 1992, in IAU Symposium: The Stellar Populations of Galaxies,
  {Barbuy}, B. and {Renzini}, A. (eds.). Dordrecht: Kluwer, Vol. 149, 225

\bibitem[{{Lattanzio} {et~al.}(1991){Lattanzio}, {Vallenari}, {Bertelli}, \&
  {Chiosi}}]{L91}
{Lattanzio}, J., {Vallenari}, A., {Bertelli}, G., \& {Chiosi}, C. 1991, A\&A,
  {\bf 250}, 340

\bibitem[{{Maeder} \& {Meynet}(1991)}]{MaederMey91}
{Maeder}, A. \& {Meynet}, G. 1991, A\&AS, {\bf 89}, 451

\bibitem[{{Mermilliod} \& {Mayor}(1989)}]{Mermi-Mayor89}
{Mermilliod}, J. \& {Mayor}, M. 1989, A\&A, {\bf 219}, 15

\bibitem[{{Meynet} {et~al.}(1994){Meynet}, {Maeder}, {Schaller}, {Schaerer}, 
\&
  {Charbonnel}}]{Meynet94}
{Meynet}, G., {Maeder}, A., {Schaller}, G., {Schaerer}, D., \& {Charbonnel}, 
C.
  1994, A\&AS, {\bf 103}, 97

\bibitem[{{Oestreicher} \& {Schmidt-Kaler}(1996)}]{Oes96}





{Oestreicher}, M. \& {Schmidt-Kaler}, T. 1996, A\&AS, {\bf 117}, 303

\bibitem[{{Panagia}(1998)}]{Panagia98}
{Panagia}, N. 1998, MmSAI, {\bf 69}, 225

\bibitem[{{Renzini}(1987)}]{Renzini87}
{Renzini}, A. 1987, A\&A, {\bf 188}, 49

\bibitem[{{Rieke} \& {Lebofsky}(1985)}]{Rieke85}
{Rieke}, G.~H. \& {Lebofsky}, M. 1985, ApJ, {\bf 288}, 618

\bibitem[{{Salasnich} {et~al.}(2000){Salasnich}, {Girardi}, {Weiss}, \&
  {Chiosi}}]{Sa00}
{Salasnich}, B., {Girardi}, L., {Weiss}, A., \& {Chiosi}, C. 2000, A\&A, {\bf
  361}, 1023

\bibitem[{{Salpeter}(1955)}]{Salp55}
{Salpeter}, E. 1955, ApJ, {\bf 121}, 161

\bibitem[{{Testa} {et~al.}(1999){Testa}, {Ferraro}, {Chieffi}, {Straniero},
  {Limongi}, \& {Fusi Pecci}}]{Te99}
{Testa}, V., {Ferraro}, F., {Chieffi}, A., {Straniero}, O., {Limongi}, M., \&
  {Fusi Pecci}, F. 1999, AJ, {\bf 118}, 2839

\bibitem[{{Unno} \& {Kondo}(1989)}]{UnnoKondo89}
{Unno}, W. \& {Kondo}, M. 1989, PASJ, {\bf 41}, 197

\bibitem[{{Vallenari} {et~al.}(1992){Vallenari}, {Chiosi}, {Bertelli},
  {Meylan}, \& {Ortolani}}]{VaC92}
{Vallenari}, A., {Chiosi}, C., {Bertelli}, G., {Meylan}, G., \& {Ortolani}, S
.
  1992, AJ, {\bf 104}, 1100

\bibitem[{{Vallenari} {et~al.}(1991){Vallenari}, {Chiosi}, {Bertelli},
  {Meylan}, \& {Ortolani}}]{V91}
{Vallenari}, A., {Chiosi}, C., {Bertelli}, G., {Meylan}, S., \& {Ortolani}, S
.
  1991, A\&AS, {\bf 87}, 517

\bibitem[{{Westerlund}(1997)}]{Westerlund97}
{Westerlund}, B. 1997, in The Magellanic Clouds (Cambridge: Cambridge
  University Press)

\bibitem[{{Xiong}(1980)}]{Xiong80}
{Xiong}, D.~R. 1980, ChA, {\bf 4}, 234

\bibitem[{{Zahn}(1991)}]{Zahn91}
{Zahn}, J. 1991, A\&A, {\bf 252}, 179

\bibitem[{{Zaritsky} {et~al.}(1997){Zaritsky}, {Harris}, \& {Thompson}}]{Z97}
{Zaritsky}, D., {Harris}, J., \& {Thompson}, I. 1997, AJ, {\bf 114}, 1933

\end{thebibliography}

\end{document}